\documentclass[11pt,a4paper]{amsart}

\usepackage{amsthm,amssymb,amsmath,epic,eepic,float}
\usepackage{rotating,epsfig,indentfirst,array,varioref}
\usepackage{appendix,marginnote,bbm,tikz,pgf,mathtools}
\usepackage{setspace,cite}

\usepackage{color}

\newtheorem{Proposition}{Proposition}[section]

\let\emptyset\varnothing

\def\ll{\left\lgroup}
\def\rr{\right\rgroup}

\def\leq{\leqslant}
\def\geq{\geqslant}

\def\pp{p^{\prime}}

  \def\rrho{{\tt R}}
\def\ssigma{{\tt C}}

\def\i{\iota}

\newdimen\tableauside\tableauside=1.0ex
\newdimen\tableaurule\tableaurule=0.4pt
\newdimen\tableaustep
\def\phantomhrule#1{\hbox{\vbox to0pt{\hrule height\tableaurule width#1\vss}}}
\def\phantomvrule#1{\vbox{\hbox to0pt{\vrule width\tableaurule height#1\hss}}}
\def\sqr{\vbox{%
  \phantomhrule\tableaustep

\hbox{\phantomvrule\tableaustep\kern\tableaustep\phantomvrule\tableaustep}%
  \hbox{\vbox{\phantomhrule\tableauside}\kern-\tableaurule}}}
\def\squares#1{\hbox{\count0=#1\noindent\loop\sqr
  \advance\count0 by-1 \ifnum\count0>0\repeat}}
\def\tableau#1{\vcenter{\offinterlineskip
  \tableaustep=\tableauside\advance\tableaustep by-\tableaurule
  \kern\normallineskip\hbox
    {\kern\normallineskip\vbox
      {\gettableau#1 0 }%
     \kern\normallineskip\kern\tableaurule}%
  \kern\normallineskip\kern\tableaurule}}
\def\gettableau#1 {\ifnum#1=0\let\next=\null\else
  \squares{#1}\let\next=\gettableau\fi\next}

\newcommand{\RR}{\mathbb{R}}

\newcommand{\ZZ}{\mathbb{Z}}

\newcommand{\bxi}{\mbox{\boldmath$\xi$}}
\newcommand{\bet}{\mbox{\boldmath$\eta$}}
\newcommand{\bxis}{\mbox{\boldmath$\scriptstyle\xi$}}
\newcommand{\bets}{\mbox{\boldmath$\scriptstyle\eta$}}

\newdimen\tableauside\tableauside=1.0ex
\newdimen\tableaurule\tableaurule=0.4pt
\newdimen\tableaustep
\def\phantomhrule#1{\hbox{\vbox to0pt{\hrule height\tableaurule
width#1\vss}}}
\def\phantomvrule#1{\vbox{\hbox to0pt{\vrule width\tableaurule
height#1\hss}}}
\def\sqr{\vbox{%
  \phantomhrule\tableaustep

\hbox{\phantomvrule\tableaustep\kern\tableaustep\phantomvrule\tableaustep}%
  \hbox{\vbox{\phantomhrule\tableauside}\kern-\tableaurule}}}
\def\squares#1{\hbox{\count0=#1\noindent\loop\sqr
  \advance\count0 by-1 \ifnum\count0>0\repeat}}
\def\tableau#1{\vcenter{\offinterlineskip
  \tableaustep=\tableauside\advance\tableaustep by-\tableaurule
  \kern\normallineskip\hbox
    {\kern\normallineskip\vbox
      {\gettableau#1 0 }%
     \kern\normallineskip\kern\tableaurule}%
  \kern\normallineskip\kern\tableaurule}}
\def\gettableau#1 {\ifnum#1=0\let\next=\null\else
  \squares{#1}\let\next=\gettableau\fi\next}

\def\be{\begin{equation}}
\def\ee{\end{equation}}
\def\ba{\begin{array}}
\def\ea{\end{array}}
\def\dps{\displaystyle}

\restylefloat{figure}

\newcommand{\cB}{\mathcal{B}}
\newcommand{\cC}{\mathcal{C}}
\newcommand{\cD}{\mathcal{D}}

\newcommand{\cF}{\mathcal{F}}

\newcommand{\cH}{\mathcal{H}}

\newcommand{\cM}{\mathcal{M}}
\newcommand{\cN}{\mathcal{N}}
\newcommand{\cO}{\mathcal{O}}

\newcommand{\cW}{\mathcal{W}}

\newcommand{\vP}{ {\vec P} }
\newcommand{\va}{ {\vec a} }
\newcommand{\vh}{ {\vec h} }
\newcommand{\vrho}{ {\vec \rho} }
\newcommand{\valpha}{ {\vec \alpha} }
\newcommand{\vomega}{ {\vec \omega} }

\newcommand{\NN}{\mathbb{N}}

\newcommand{\apos}{\alpha_{+}}
\newcommand{\aneg}{\alpha_{-}}

\newtheorem{ca}{Figure}

\def\ll{ \left\lgroup}
\def\rr{\right\rgroup}

\newcommand{\checkedcell}{{\makebox[0pt][l]{$\square$}\raisebox{.15ex}{\hspace{0.1em}$\checkmark$}}}

\begin{document}

\title[AGT, $N$-Burge partitions and $\cW_N$]{
AGT, $N$-Burge partitions and $\cW_N$ minimal models}

\author[]{
Vladimir Belavin
\!\!$^{{\scriptstyle \mathbbm{\, A}}}$
, 
Omar Foda 
\!\!$^{{\scriptstyle \mathbbm{\, B}}}$
and 
Raoul Santachiara
\!\!$^{{\scriptstyle \mathbbm{\, C}}}$
}

\address{
\!\!\!\!\!\!\!\!\!
$^{{\scriptstyle \mathbbm{A}}}$
I E Tamm Department of Theoretical Physics,
P N Lebedev Physical Institute,
Leninsky Avenue 53, 119991 Moscow, Russia, 
\newline
Department of Quantum Physics,
Institute for Information Transmission Problems,
Bolshoy Karetny per. 19, 127994 Moscow, Russia
\newline
$^{{\scriptstyle \mathbbm{B}}}$
Mathematics and Statistics, University of Melbourne, Parkville, VIC 3010, Australia
\newline
$^{{\scriptstyle \mathbbm{C}}}$
Laboratoire de Physique Th\'eorique et Mod\`eles Statistiques,
CNRS UMR 8626, Bat. 100, Universit\'e Paris-Sud, 91405 Orsay cedex, France
}

\email{
vlbelavin@yandex.ru,
omar.foda@unimelb.edu.au, 
raoul.santachiara@gmail.com
}

\keywords{
$\cW_N$ AGT. 
Burge partitions.
$\cW_N$ minimal models.
}

\begin{abstract}
Let $\cB^{\, p, \, \pp, \, \cH}_{N, n}$ be a conformal block, with $n$ consecutive channels 
$\chi_{\i}$, $\i = 1, \cdots, n$, in the conformal field theory $\cM^{\, p, \, \pp}_N \! 
\times \! \cM^{\cH}$, where $\cM^{\, p, \, \pp}_N$ is a $\cW_N$ minimal model, generated
by chiral spin-2, $\cdots$, spin-$N$ currents, and labeled by two co-prime integers $p$ 
and $\pp$, $1 < p < \pp$, while $\cM^{\cH}$ is a free boson conformal field theory. 
$\cB^{\, p, \, \pp, \cH}_{N, n}$ is the expectation value of vertex operators between 
an initial and a final state. Each vertex operator is labelled by a charge vector that 
lives in the weight lattice of the Lie algebra $A_{N-1}$, spanned by weight vectors 
$\vomega_1, \cdots, \vomega_{N-1}$. We restrict our attention to conformal blocks with 
vertex operators whose charge vectors point along $\vomega_1$. The charge vectors that 
label the initial and final states can point in any direction.

Following the $\cW_N$ AGT correspondence, and using Nekrasov's instanton partition functions 
without modification to compute $\cB^{\, p, \, \pp, \cH}_{N, n}$, leads to ill-defined expressions. 
We show that restricting the states that flow in the channels $\chi_{\i}$, $\i = 1, \cdots, n$, 
to states labeled by $N$ partitions that we call $N$-Burge partitions, that satisfy conditions 
that we call $N$-Burge conditions, leads to well-defined expressions that we propose to identify 
with $\cB^{\, p, \, \pp, \, \cH}_{N, n}$. 
We check our identification by showing that a non-trivial conformal block that we 
compute, using the $N$-Burge conditions satisfies the expected differential equation. Further, 
we check that the generating functions of triples of Young diagrams that obey 3-Burge conditions
coincide with characters of degenerate $\cW_3$ irreducible highest weight representations.
\end{abstract}

\maketitle

%SECTION.01
\section{Introduction}
\label{introduction}

{\it We propose a modification of the $\cW_N$ AGT correspondence so that it applies to $\cW_N$ 
minimal models, and use it compute $\cW_N$ minimal model conformal blocks that are expectation 
values of vertex operators whose charge vectors are vectors in the $A_{N-1}$ weight lattice, 
that point along the direction of the fundamental weight vector $\vomega_1$.}

\subsection{The AGT correspondence} The original AGT correspondence, or simply AGT, is the statement 
that the instanton partition functions of 4D linear and cyclic $U(2)$ quiver $\cN \! = \! 2$ 
supersymmetric gauge theories are equal, up to a Heisenberg factor, to Virasoro conformal blocks on 
the sphere and on the torus, respectively, with non-minimal central charges \cite{agt}. 
It was conjectured by Alday, Gaiotto and Tachikawa in \cite{agt}, proven in important special cases 
in \cite{fateev.litvinov.2009, mironov.morozov.shakirov.01, mironov.morozov.shakirov.02, 
hadasz.jaskolski.suchanek.2010, schiffmann.vasserot}, then proven in full generality by Alba, Fateev, 
Litvinov and Tarnopolskiy in \cite{alba}. 

\subsection{The $\cW_N$ AGT correspondence}
The correspondence was extended to an identification, also up to a Heisenberg factor, of the instanton 
partition functions of 4D linear and cyclic $U(N)$ quiver $\cN \! = \! 2$ supersymmetric gauge theories 
and conformal blocks in $\cW_N$ conformal field theories
\footnote{\, 
We take $\cW_N$ to be the infinite-dimensional algebra generated by chiral spin-2, $\cdots$, spin-$N$
currents, referred to as $\cW (2, 3, \cdots, N)$ in \cite{bouwknegt.schoutens.review}.  
}
on the sphere and on the torus, with non-minimal central charges \cite{wyllard, mironov.morozov, 
fateev.litvinov.2011}. However, the latter identification is restricted to a class of $\cW_N$ 
conformal blocks, with non-minimal central charges, characterised by a condition discussed by 
Fateev and Litvinov \cite{fateev.litvinov.2008} and by Wyllard \cite{wyllard}. 

\subsection{The condition of Fateev, Litvinov and Wyllard} 
\label{flw.condition}
Consider a $\cW_N$ Toda conformal block that consists of $n$ consecutive channels, that is, the 
expectation value of $(n+3)$ $\cW_N$ vertex operators. Each vertex operator represents a $\cW_N$ 
highest weight state that is labelled by an $(N-1)$-component charge-vector that lives in the 
$A_{N-1}$ weight lattice spanned by the fundamental weight vectors $\{\vomega_1, \cdots, \vomega_{N-1}\}$. 

The $\cW_N$ AGT correspondence applies to $n$-channel $\cW_N$ conformal blocks that involve $(n+3)$ 
vertex operators that consist of a $\cW_N$ factor and a Heisenberg factor, such that the $A_{N-1}$ 
charge-vector of two of these operators can point in any direction along the $A_{N-1}$ weight lattice, 
while the charge vectors of the remaining $(n+1)$ operators are restricted to point along the same 
direction, for example $\vomega_1$, or directions that are related to $\vomega_1$ by Weyl reflections. 
In the sequel, we refer to this condition as the FLW condition.

\subsection{$\cW_N$ AGT in non-minimal $\cW_N$ models} Applying the AGT prescription to $\cW_N$ 
conformal blocks, that is, identifying $U(N)$ instanton partition functions with conformal blocks, 
up to Heisenberg factors, makes perfect sense for conformal field theories with non-minimal central 
charges. 

The instanton partition functions are in the form of sums over products of rational functions of 
the parameters of the theory, as in (\ref{agt.gen}). Each sum corresponds to a gauge group in the 
quiver gauge theory. The terms in a sum are parameterised by the set of all possible $N$-partitions. 
There are no conditions on the partitions that we are allowed to sum over.

On the conformal field side of the AGT correspondence, each term in a sum corresponds to a state 
in a Verma module of the algebra $\cW_N \! \times \! \cH$, where $\cH$ is the Heisenberg algebra. 
The fact that we sum 
over all possible $N$ partitions corresponds to allowing all possible states that live in a $\cW_N$ 
Verma module, times a Heisenberg module, to flow in the channels of the conformal block. 

\subsection{$\cW_N$ AGT in minimal $\cW_N$ models}
Choosing the parameters of the instanton partition functions such that one obtains minimal models 
on the conformal field theory side of AGT leads to zeros in the denominators of the instanton 
partition functions. These singularities are non-physical and can be traced to the fact that by 
summing over all possible 
states in the $\cW_N$ modules that flow through the channels of the conformal blocks, one allows for 
the flow of null states that should decouple. 

One approach to remove these non-physical singularities is to enforce the fusion rules at the level
of the instanton partition functions. This requires that we analytically continue the instanton 
partition functions, in a way that preserves the fusion rules, then show that to each zero in 
the denominator of a summand, there is a higher order zero in the numerator of the same factor, such 
that the corresponding null state decouples in the appropriate limit.

\subsection{Restricting the Young diagrams} 
In this work, we choose to follow a different approach. That is, we characterise the sets of $N$ 
partitions that lead to null states and exclude them from the sums. This is the approach that
was followed in \cite{alkalaev.belavin, bershtein.foda} to obtain conformal blocks in Virasoro 
minimal models. In \cite{alkalaev.belavin, bershtein.foda}, this procedure led to well-defined 
expressions. The proposal of \cite{alkalaev.belavin, bershtein.foda} is that these expressions 
can be identified with minimal model conformal blocks, up to Heisenberg factors, which was 
checked to be the case in a number of non-trivial examples. The present work is an extension 
of the proposal of \cite{alkalaev.belavin, bershtein.foda} to $U(N)$ instanton partition functions 
and $\cW_N$ conformal blocks. 

\subsection{$N$-Burge partitions} 
\label{restricted.N.partitions}
Our goal is to provide AGT relations  for $\cB^{\, p, \, \pp, \cH}_{N, n}$. The difficulty is 
that naively applying (\ref{agt.gen}) to minimal $\cW_N$ model, one gets singular expressions, 
as explained in detail in the context of $\cW_2$ in \cite{alkalaev.belavin, bershtein.foda}. 
The origin of these singularities is related to the existence of zero-norm states in the $\cW_N$ 
Verma modules with central charge  (\ref{c.p.p.prime}) and associated to the vectors 
(\ref{charge.vector}). Summation in (\ref{agt.gen}) includes states in a Verma module rather 
than in an irrep of $\cW_N^{\, p, \, \pp} \, \times \, \cH$ and therefore containing the 
contribution of zero-norm states.  These states should be removed when computing minimal model 
conformal blocks. 

In this work, as in \cite{alkalaev.belavin, bershtein.foda}, we avoid these zeros by restricting 
the summations over the $N$-partitions that appear in the sum (\ref{agt.gen}). We provide 
an expression of $\cB^{\, p, \, \pp, \cH}_{N, n}$ in terms of a sum of the type (\ref{agt.gen}) 
that consists of products of factors $Z^{\i}_{bb}$. Each $Z^{\i}_{bb}$ is an expectation value 
of an $\cW_N \! \times \! \cH$ vertex operator $\cO_{\i}$, characterised by a charge vector 
$\va_{m_{\i} n_{\i}}$. This charge vector lives in the $A_2$ weight lattice and points in the 
direction of the fundamental weight $\vomega_1$. 
The expectation value of $\cO_{\i}$ is computed between $\cW_N \! \times \! \cH$ basis states, 
an  in-state labeled by a charge vector $\vP_{\vec r_{\i - 1} \ \vec s_{\i - 1}}$, 
and $N$ partitions $\vec Y^{\i - 1}$, and
an out-state labeled by a charge vector $\vP_{\vec r_{\i    } \ \vec s_{\i    }}$, 
and $N$ partitions $\vec Y^{\i    }$.

The charge vectors $\vP_{\vec r_{\i - 1} \ \vec s_{\i - 1}}$ and $\vP_{\vec r_{\i    } \ \vec s_{\i    }}$
are chosen such that degenerate $\cW_N \! \times \! \cH$ highest weight modules flow in the intermediate 
channels. Given this choice, $Z^{\i}_{bb}$ which is a rational function of its parameters, can have zeros 
in the denominators, leading to ill-defined expressions. We characterize the singularities in $Z^{\i}_{bb}$ 
that lead to ill-defined expressions, and attribute these singularities to zero-norm states that should not 
be allowed to propagate in the channels of the minimal model conformal blocks. We eliminate these zero-norm 
states by restricting the $N$-partitions that label the states that flow in the $\i$-th channel in a minimal 
model conformal block, and that appear in (\ref{agt.gen}) to $N$-partitions ${\vec Y}=\{Y_1, \cdots, Y_N\}$, 
that satisfy the conditions

\begin{equation}
\label{N.Burge.conditions}
\boxed{
Y_{i, \, \rrho}-Y_{i+1, \, \rrho+s_i-1}\geq -r_i+1
}
\end{equation}

\noindent where $Y_{i, \, \rrho}$ is the $\rrho$-row of $Y_i$, $i = 1, \cdots, N$, $r_i$ and 
$s_i$, $i = 1, \cdots, N$, $\sum_{i=1}^N r_i = p$, $\sum_{i=1}^N s_i = \pp$,
are parameters that characterise the $\cW_N$ irreducible highest weight module that flows in 
the $\i$-th channel under consideration, and $Y_{N+1} = Y_1$
\footnote{\,
For explicit examples of partitions that satisfy $N$-Burge conditions, see subsection {\bf \ref{checking}}.
}

These restricted {\it $N$-Burge partitions} were introduced, in case $N=2$, in \cite{burge},
further studied in \cite{foda.lee.welsh} and appeared in full generality in 
\cite{gessel.krattenthaler, feigin.feigin.jimbo.miwa.mukhin.02}. We show that when used 
to restrict AGT to compute $\cB^{\, p, \, \pp, \, \cH}_{N, n}$, that is when we sum over $N$-Burge 
partitions rather than on all possible $N$-partitions, we obtain
\begin{equation}
\label{restricted.sum}
\boxed{
\cB^{\, p, \, \pp, \, \cH}_{N, n} =\sum^{\prime}_{\vec Y^1, \cdots, \vec Y^n}
\prod_{\i = 1}^{n+1} q_{\i}^{| \vec Y^{\i } |} 
Z_{bb}^{\i} 
\ll \vP_{\vec r_{\i - 1} \ \vec s_{\i - 1}}, \vec Y^{\i - 1}\ | \ \va_{m_{\i} n_{\i} } \ | 
  \ \vP_{\vec r_{\i    } \ \vec s_{\i    }}, \vec Y^{\i} \rr
}
\end{equation}
\noindent where $\sum^{\prime}$ indicates that the sum is restricted to $N$-partitions that 
satisfy the $N$-Burge conditions (\ref{N.Burge.conditions}), we obtain well-defined expressions. 
Brief explanations of the meaning of the various terms in equation (\ref{restricted.sum}) were 
given in earlier paragraphs. More details definitions can be found in section {\bf \ref{wn.algebras}}.

\subsection{Outline of contents}
\label{outline.of.contents}
%section.02
In section {\bf \ref{wn.algebras}}, we recall the basics of $\cW_N$ algebras and conformal field 
theories, and in
%section.03
{\bf \ref{wn.agt}}, 
we do the same for the original, unmodified $\cW_N$ AGT correspondence.
%section.04
In {\bf \ref{agt.minimal}}, we discuss the restrictions that we need to impose on the
$N$-partitions that are summed over in Nekrasov's instanton partition functions in order 
to compute conformal blocks in minimal $\cW_N$ models.
%section.05
In {\bf \ref{w.3.minimal.models}}, we recall basic facts related to the $\cW_3$ minimal models,
then we check the $N$-Burge conditions obtained in {\bf \ref{agt.minimal}}, in the context 
of $\cW_3$ minimal models, by considering conformal blocks that satisfy differential equations. 
%section.06
In {\bf \ref{wn.characters}}, we discuss the derivation of the characters of $\cW_3$ minimal 
models from the 3-Burge partitions.
%section.07
Finally section {\bf \ref{remarks}} contains a number of remarks.

%\vfill
%\newpage

%SECTION.02
\section{$\cW_N$ algebras and conformal field theories}
\label{wn.algebras}

{\it We recall basic definitions related to $\cW_N$ algebras, 
               with focus on $\cW_3$, followed by 
               basic definitions related to $\cW_N$ conformal field theories, 
	       with focus on $\cW_3$ minimal models.} 

\subsection{$\cW_N$ algebras} The Virasoro algebra is generated by the Laurent components of the 
holomorphic part of stress-energy tensor $T(z)$ which is a spin-2 chiral field 
\cite{belavin.polyakov.zamolodchikov}. 
The $\cW$ algebras are extensions of the Virasoro algebra, generated by higher-spin chiral fields. 
For a comprehensive review, see \cite{bouwknegt.schoutens.review}. In this work, we use $\cW_N$ 
to indicate the infinite-dimensional algebra generated by chiral fields of spin $2, 3, \cdots, N$, 
referred to as $\cW (2, 3, \cdots, N)$ in \cite{bouwknegt.schoutens.review}.

\subsubsection{The $\cW_3$ algebra} 
The $\cW_N$ algebra, for large $N$, is a complicated object. To be concrete, we choose to work in 
terms of examples from $\cW_3$, which is the simplest $\cW_N$ algebra beyond Virasoro, and that 
has the basic features of higher $N$ $\cW_N$ algebras, particularly the fact that it is not a Lie
algebra.

$\cW_3$ is generated by 
the Laurent components of the chiral spin-2 stress-energy tensor $T(z)$,   $L_n$, $n \in \ZZ$, and 
the Laurent components of a   chiral spin-3 current              $\cW(z)$, $W_n$, $n \in \ZZ$ 
\cite{zamolodchikov.w.algebra}. 
Following the notation and conventions of \cite{iles.watts}, the defining relations of the $\cW_3$ 
algebra are

\begin{align}
\label{w3.algebra}
&\left[ L_m, L_n \right] = (m-n) \, L_{m+n} + \frac{c}{12} (m^3-m) \delta_{m+n, 0},
\\
\nonumber
&\left[ L_m, W_n \right] = (2m-n) \, W_{m+n}, 
\\
\nonumber
&\left[ W_m, W_n \right] = 
(m-n) \ll \frac{1}{15} (m+n+3) (m+n+2) - \frac{1}{6} (m+2) (n+2) \rr \, L_{m+n}
\\
\nonumber
&\hspace{40mm}+\ll \frac{c}{360} \rr m (m^2 - 4) (m^2 - 1) \delta_{m+n, 0} 
+ \beta \, (m-n) \, \Lambda_{m+n},  
\end{align}

\noindent where 

\begin{equation}
\label{beta}
\beta = \frac{16}{22 + 5c}, \quad
\Lambda_m = \sum_{p \geq -1} L_{m-p} \, L_{p  }
          + \sum_{p \leq -2} L_{p  } \, L_{m-p}
	  - \frac{3}{10}(m+2)(m+3) L_m,
\end{equation}
\noindent and $c$ is the Virasoro central charge. 

\subsubsection{$\cW_3$ heighest weight states and eigenvalues}
Consider a vector $\vP$ in the weight lattice of the Lie algebra $A_2$, spanned by 
the fundamental weight vectors $\vomega_1$ and $\vomega_2$,

\begin{equation}
\vP = P_1 \, \vomega_1 + P_2 \, \vomega_2,
\end{equation}

\noindent where $P_1, P_2 \in \RR$. The $\cW_3$ highest weight state $|\vP \, \rangle$, 
associated to $\vP$, is defined by  

\be 
 L_0 | \vP \rangle = \Delta_{\vP} \;| \vP \rangle,\quad
 W_0 | \vP \rangle =      w_{\vP} \;| \vP \rangle
\ee

\noindent The eigenvalues $\Delta_{\vP}$ and $w_{\vP}$ are given in terms of $\vP$ by 

\be 
\label{conformal.dimension}
\Delta_{\vP}  = \frac{c-2}{12} + \frac12 {\vP}^2, \quad 
     w_{\vP}  = \sqrt{-3 \beta} \ \prod_{i=1}^3 \langle \vP \, | \, \vh_i \rangle,
\ee

\noindent where $\vh_i$, $i = 1, 2, 3$, are the weight vectors of the first fundamental 
representation of the Lie algebra $A_2$, and $\langle \vP \, | \, \vh_i \rangle$ is a scalar 
product on the $A_2$ weight lattice. The vertex operator $\cO_{\vP}$ is associated to the 
$\cW_3$ highest weight state $| {\vP} \rangle$. 

Another standard parametrisation of the highest weight vectors, and the corresponding $\cW_3$ 
vertex operators, is in terms of the vector charge $\va_{\vP}$, 

\be 
\va_{\vP} = \ll b + b^{-1}\rr \ll \vomega_1 + \vomega_2 \rr  + \vP
\label{vertexcharge}
\ee

\noindent In the above Toda-like notation, $b$ parametrizes the central charge

\begin{equation}
\label{central.charge}
c = 2 + 24 \ll b+ b^{-1}\rr^2
\ee   

\subsubsection{Higher rank $\cW_N$ algebras}
The $\cW_4$, or $\cW(2, 3, 4)$ algebra generated by spin 2, 3, and 4 chiral fields is defined in 
\cite{blumenhagen.1991, kausch.watts.1991}. The definition of the algebras for higher $N$ 
is involved. We refer to \cite{bouwknegt.schoutens.review} for a complete discussion. 

\subsection{$\cW_N$ conformal field theories} 
We take
$\cM_N \! \times \! \cM^{\cH}$ to be the 2-dimensional conformal field theory that 
consists of a factor $\cM_N$ based on the infinite-dimensional $\cW_N$ algebra, and 
a factor $\cM^{\cH}$ based on the Heisenberg algebra $\cH$. 
In this work, we focus on minimal models, and write $\cM_N = \cM_N^{p, \, \pp}$. 

\subsubsection{The labels of the minimal $\cW_N$ models, the background charge parameter and 
the screening charge parameters} 
A minimal $\cW_N$ model, $\cM^{\, p, \pp}_N$, is labeled by two coprime integers, $p$ and $\pp$, 
$1 < p < \pp$. More precisely, the central charge $c^{\, p, \, \pp}_N$, of $\cM^{\, p, \pp}_N$,
is

\begin{equation}
\label{c.p.p.prime}
c_N^{\, p, \, \pp} = (N-1) \ll 1 - N (N+1) \, \alpha_0^2 \rr, 
\end{equation}

\noindent where $\alpha_0$ is the background charge parameter

\begin{equation}
\label{background.charge}
\alpha_0 = \apos + \aneg,
\end{equation}

\noindent and $\apos$ and $\aneg$ are the screening charge parameters

\begin{equation}
\label{screening.charges}
\apos =    \ll \frac{\pp}{p} \rr^{\frac12},  
\quad 
\aneg =  - \ll \frac{p}{\pp} \rr^{\frac12} 
\end{equation}

\subsubsection{Remarks} 
{\bf 1.} The definition (\ref{background.charge}) of the background charge parameter $\alpha_0$, as 
well as the definition (\ref{charge.vector}) of the $A_{N-1}$ charge vector, below, are 
appropriate for minimal models based on $\cW_N$ algebras, such that $N \geq 3$, but differ 
from the definitions used in minimal models based on the Virasoro algebra $\cW_2$. In this 
work, we focus on $\cW_N$ minimal models, such that $N \geq 3$.
\noindent {\bf 2.} The $\cW_N$ AGT correspondence is discussed in \cite{mironov.morozov, 
wyllard} in the context of non-minimal $\cW_N$ models. These models are not labeled by 
coprime integers, and their central charges do not satisfy (\ref{c.p.p.prime}).

\subsubsection{Minimal $\cW_N$ conformal blocks $\cB_{N, n}^{\, p, \pp}$} We are interested 
in computing linear conformal blocks $\cB_N^{\, p, \pp}$ in $\cW_N$ minimal models 
$\cM_N^{\, p, \pp}$ that can be represented schematically as in Figure {\bf \ref{the.comb.diagram}}.

%FIG.01
\begin{center}
\begin{minipage}{5.0in}
\setlength{\unitlength}{0.1cm}
\renewcommand{\dashlinestretch}{30}
\begin{picture}(40, 25)(-40, 00)
%
% negative x shifts the figure to the right
% positive x shifts the figure to the left
% negative y shifts the figure up
% positive y shifts the figure down
\thicklines
%%%%%%%%%%%%%%%%%%%%%%%
%Two horizontal segments
\put(00,00){\line(1,0){25}}
\put(34,00){\line(1,0){25}}
%%%%%%%%%%%%%%%%%%%%%%%
%Four vertical segments
\put(10,00){\line(0,1){10}}
\put(20,00){\line(0,1){10}}
\put(40,00){\line(0,1){10}}   
\put(50,00){\line(0,1){10}}   
%%%%%%%%%%%%%%%%%%%%%%%
%
\put( 07,12.0){\mbox{$\cO_1$}}
\put( 17,12.0){\mbox{$\cO_2$}}
\put( 38,12.0){\mbox{$\cO_n$}}
\put( 48,12.0){\mbox{$\cO_{n+1}$}}
\put(-01,-5.5){\mbox{$\cO_0$}}
\put( 54,-6.0){\mbox{$\cO_{n+2}$}}
\put( 12,-6.0){\mbox{$\chi_1$}}
\put( 41,-6.0){\mbox{$\chi_n$}}
%%%%%%%%%%%%%%%%%%%%%%%
\put(27,10){\mbox{$\cdots$}}
\put(27,00){\mbox{$\cdots$}}
%%%%%%%%%%%%%%%%%%%%%%%%
\end{picture}
\bigskip
\bigskip
\bigskip
\begin{ca}
\label{the.comb.diagram}
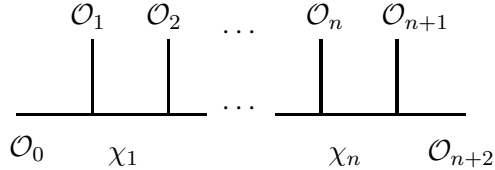
The comb diagram of an $n$-channel linear conformal block. The initial and final states, $\cO_0$ and $\cO_{n+2}$, 
the vertex operators $\cO_i$, $i=1, \cdots, n+1$, and the $\cW_N$ irreducible highest weight representation that 
flows in the $\i$-th channel $\chi_{\i}$, $\i = 1, \cdots, n$, are defined in the text. 
\end{ca}
\end{minipage}
\end{center}
\bigskip

An $n$-channel conformal block $\cB_{N, n}^{\, p, \pp}$ is an expectation value of $(n+2)$ $\cW_N$ 
vertex operators, $\cO_i$, $i=1, \cdots, n+1$, inserted at different points on a Riemann surface, which 
in this work we take to be a Riemann sphere
\footnote{\, 
Except in section {\bf \ref{wn.characters}}, when we discuss the characters of degenerate $\cW_3$ 
irreducible highest weight representations, which are essentially 0-point conformal blocks on the 
torus.}, 
between an initial state and a final state, such that the states that flow in channel $\chi_\i$,
$\i = 1, \cdots, n$, between two consecutive insertions, belong to one and only one $\cW_N$ irreducible 
highest weight representation, $\cH_{\i}$. In the following, we specify the parameters that label 
the vertex operators $\cO_i$ and the highest weight representations $\chi_\i$.

\subsubsection{Labels of vertex operators} 
\label{primary.fields}
A vertex operator $\cO_i$ of $\cM^{\, p, \, \pp}_N$ is labelled by two 
sets of non-zero positive integers 
$\vec r = \{r_1, \cdots, r_{N-1} \}$  and 
$\vec s = \{s_1, \cdots, s_{N-1} \}$, that satisfy 

\begin{equation}
1 \leq  \ll \sum_{i=1}^{N-1} r_i \rr \leq  p, \quad 
1 \leq  \ll \sum_{i=1}^{N-1} s_i \rr \leq \pp
\end{equation}
 
\noindent It is useful to define two more non-zero positive integers $r_N$ and $s_N$, such that

\begin{equation}
\label{sum.conditions}
\sum_{i=1}^N r_i =   p, \quad
\sum_{i=1}^N s_i = \pp
\end{equation} 

\subsubsection{Charge vectors of vertex operators} 
The vertex operators $\cO_i$,   $i = 1, \cdots, n+1$, are represented by external vertical line 
segments in Figure {\bf \ref{the.comb.diagram}}. 
Each $\cO_i (z_i)$ is labelled by a vector charge $a_{\, \vec r_i \, \vec s_i}$, that has 
$(N-1)$ components, parameterised in terms of the screening charge parameters as
\begin{equation}
\label{vector.charge}
\va_{\, \vec r_i \, \vec s_i} = 
\sum_{i=1}^N \ll \ll 1 - r_i \rr \apos + \ll 1 - s_i \rr \aneg \rr \vomega_i,
\end{equation}

\noindent where the parameters $\vec r$ and $\vec s$ were discussed in paragraph {\bf \ref{primary.fields}}
\footnote{\,
The indices $i$ in $\vec r_i$   and $\vec s_i$ on the left hand side of (\ref{vector.charge}) 
refer to the corresponding vertex operator and are not summed over. 
The indices $i$ in $r_i$, $s_i$ and $\vomega_i$ on the right hand side 
refer to the fundamental weight vectors of the $A_{N-1}$ and are summed over.
The integers $r_i$ and $s_i$ on the right hand side are the components of the vectors 
$\vec r_i$ and $\vec s_i$ on the left hand side, respectively.
}. 
We include these details by writing $\cO_i$ as $\cO_{\, \vec r_i \, \vec s_i} (z_i)$.
The initial and final states correspond to the vertex operators $\cO_0 (z_0)$ and $\cO_{n+2} (z_{n+2})$.
In this work, we take the charge vectors of $\cO_0 (z_0)$ and $\cO_{n+2} (z_{n+2})$ to be arbitrary, 
the charge vectors of all remaining vertex operators to satisfy the FLW condition, and we write
\begin{equation}
\va_{\, \vec r_i \, \vec s_i} = 
a_{\,      r_1         s_1} \vomega_1 = 
\ll \ll 1 - r_1 \rr \apos + \ll 1 - s_1 \rr \aneg \rr \, \vomega_1, 
\quad i = 1, \cdots, n+1
\end{equation}

\subsubsection{Charges of the highest weight states that flow in the channels} 
The channels $\chi_\i$, $\i = 1, \cdots, n$, are represented by internal line segments in Figure 
{\bf \ref{the.comb.diagram}}. 
In $\cW_N$ minimal models, each channel $\chi_\i$ carries states that belong to a degenerate $\cW_N$ 
irreducible highest weight representation $\cH^{\, p, \pp}_{\, \vec r_\i \, \vec s_\i}$. 
Each of these representations consists of a highest weight state and infinitely-many descendents. 
The highest weight state of $\cH^{\, p, \pp}_{\, \vec r_\i \, \vec s_\i}$ is labelled by a charge 
that flows between a vertex operator at $z_i$ and a vertex operator at $z_{i+1}$, $i = 1, \cdots, n$. 

\subsubsection{The charge vectors} 
The $A_{N-1}$ charge vector $\vP_{\, \vec r  \, \vec s}$, of a degenerate $\cW_N$ irreducible
highest weight representation $\cH^{\, p, \pp}_{\, \vec r \, \vec s}$, is defined as

\begin{equation}
\label{charge.vector}
\vP_{\, \vec r  \, \vec s} = - \sum_{i=1}^{N-1} \ll r_i \apos + 
                                                  s_i \aneg \rr \vomega_i,
\end{equation}

\noindent where $\vomega_i$, $i = 1, \cdots, N$, are the $A_{N-1}$ fundamental weight vectors. 

\subsubsection{The conformal dimensions}
The conformal dimension $\Delta_{\, \vec r \, \vec s}$ of the vertex operator $\cO_{\, \vec r \, \vec s}$ 
that carries a charge vector $\vP_{\, \vec r  \, \vec s}$ is
\begin{equation} 
\label{hrs}
\Delta_{\, \vec r \, \vec s} = \frac12 \ll \vP_{\, \vec r \, \vec s} + \alpha_0 \, \vrho \rr 
                                        {\bf \cdot}
                                  \ll \vP_{\,   \vec r \, \vec s} - \alpha_0   \, \vrho   \rr
                        = \frac12 \ll \vP^2_{\, \vec r \, \vec s} - \alpha_0^2 \, \vrho^2 \rr
\end{equation}

\noindent where the product in the middle term of equation (\ref{hrs}) is a scalar product 
of two vectors, $\vP^2$ and $\vrho^2$ are the squares of the norms of the charge vector $\vP$ 
and the Weyl vector $\vrho$, 

\begin{equation}
\vrho = \sum_{i=0}^{n-1} \vomega_i
\end{equation}

\subsubsection{The degenerate irreducible highest weight representations of $\cM^{\, p, \, \pp}_N$} 
These are obtained from the corresponding Verma modules by factoring out the submodules 
that consist of zero-norm states and their descendants. It can be shown that in the 
representation associated to $\cO_{\, \vec r \, \vec s}$ there are $(N-1)$ zero-norm 
states of conformal dimensions $\Delta_{\, \vec r \, \vec s} + r_i s_i$, $i=1, \cdots, N-1$.

Following \cite{agt, wyllard}, we introduce an auxiliary free boson theory $\cM^{\, \cH}$, 
compute conformal blocks $\cB_{N, n}^{\, p, \pp, \cH}$ in $\cM_N^{\, p, \pp} \times \cM^{\, \cH}$, then 
factor out the Heisenberg contribution $\cM^{\, \cH}$, which is known. Before we do that, we 
need to recall basic definitions related to $\cB_{N, n}^{\, p, \pp}$. 

\subsubsection{$\cM^{\, p, \, \pp}_N$ vertex operators}
\label{first.factor.vertex.operator}
A vertex operator $\cO_{\, \vec r \, \vec s}$, in $\cM^{\, p, \, \pp}_N$, located at $z$ on the 
Riemann sphere, is represented, at operator level, as a vertex operator. In the Coulomb gas 
representation of $\cW_N$ minimal models, a vertex operator is represented as an exponential 
of $(N-1)$ free bosons, $\phi_i, i = 1, \cdots, N-1$, that live in the $A_{N-1}$ root lattice,

\begin{equation}
\label{minimal.model.vertex.operator}
\cO_{\, \vec r, \vec s} (z) = e^{i \, \vec a \, \cdot \, \vec \phi (z)}, \quad
\vec a        = \sum_{i=1}^{N-1}    a_i    \, \vomega_i,                  \quad
\vec \phi (z) = \sum_{i=1}^{N-1} \phi_i (z)  \, \valpha_i, 
\end{equation}

\noindent where $\valpha_i$ and $\vomega_i$, $i=1, \cdots, N-1$, are the fundamental root and 
weight vectors of $A_{N-1}$. In this work, we focus on vertex operators that satisfy the FLW 
condition discussed in subsection {\bf \ref{flw.condition}}, that is 

\begin{equation}
\label{charge.01}
\va_{\, \textit{FLW}} = a_1 \vomega_1
\end{equation}

\subsubsection{$\cM^{\, \cH}$ vertex operators}
\label{second.factor.vertex.operator}
As we will see in the sequel, a modification of the $\cW_N$ AGT prescription, obtained by 
restricting the Young diagrams that we sum over, will provide us with well-defined expressions 
that we identify with $\cW_N$ conformal blocks that satisfy the FLW condition,
in $\cM^{\, p, \, \pp}_N \! \times \! \cM^{\, \cH}$ conformal field theories. These conformal 
blocks are expectation values of holomorphic vertex operators that consist of two factors. One 
factor belongs to $\cM^{\, p, \, \pp}_N$ and was discussed in paragraph 
{\bf \ref{first.factor.vertex.operator}}. The other factor belongs to $\cM^{\, \cH}$ and has,
at this stage, the form
\footnote{\,
As we will see shortly, in the case that we are interested in, the charge $\alpha_N$ of the 
$\cM^{\, \cH}$ vertex operator is completely fixed by the charge of the $\cM^{\, p, \, \pp}_N$
vertex operator that will multiply it, at the same point on the Riemann sphere.
}

\begin{equation}
\label{carlsson.okounkov}
\cO^{\, \cH} (z) = e^{ i \ll \alpha_0 - \alpha_N \rr \phi_N^+ (z)} 
                   e^{ i                \alpha_N     \phi_N^- (z)}, 
\end{equation}

\noindent where $\phi_N^+$ and $\phi_N^-$ are the positive frequency and negative frequency components of 
the holomorphic factor of an $N$-th, independent free boson $\phi_N$, and the charges of the exponentials 
of these components are chosen to different, as in equation (\ref{carlsson.okounkov}), where $\alpha_0$ 
is the background charge parameter. This vertex operator, in which no zero-mode appears, first appeared in 
\cite{carlsson.okounkov} and was studied further in \cite{alba}. 

\subsubsection{$\cM^{\, p, \, \pp}_N \! \times \! \cM^{\, \cH}$ conformal blocks} 
A conformal block is an expectation value of holomorphic vertex operators. We use 
$\cB^{\, p, \, \pp}_{N, n}$, $\cB_n^{\, \cH}$ and $\cB^{\, p, \, \pp, \, \cH}_{N, n}$, 
for a linear conformal block, with $n$ consecutive channels in 
$\cM^{\, p, \, \pp}_N$, $\cM^{\, \cH}$, and $\cM^{\, p, \, \pp}_N \! \times \! \cM^{\, \cH}$,
respectively. Only conformal blocks that live on the Riemann sphere, with $n$ consecutive 
channels, as in Figure {\bf \ref{the.comb.diagram}}, are considered in this work. The extension 
to cyclic conformal blocks 
on the torus is straightforward by allowing the initial and final states to be descendants, 
identifying them, then summing over all possible descendants. 

Our notation is such that an 
$n$-channel conformal block $\cB^{\, \textit{indices}}_{N, n}$, is the expectation value of 
$(n+3)$ chiral vertex operators
$\cO^{\, \textit{same indices}}_{\i} (z_{\i})$, ${\i} = 0, \cdots, (n+2)$, in 
$\cM^{\, \textit{same indices}}$ and $z_{\i}$ are the coordinates of the vertex insertions.

We wish to compute $\cB^{\, p, \, \pp, \, \cH}_{N, n}$. In this case, each vertex operator 
is a product of two vertex operators, 
$\cO^{\, p, \, \pp}_{\i} (z_{\i}) \! \times \! \cO^{\cH} (z_{\i})$, where 
$\cO^{\, p, \, \pp}_{\i} (z_{\i})$ is in $\cM^{\, p, \, \pp}_N$, 
$\cO^{\cH}               (z_{\i})$ is in $\cM^{\, \cH}$, 
and the charge of $\cO^{\cH} (z_{\i})$ is completely determined by that of 
$\cO^{\, p, \, \pp}_{\i} (z_{\i})$, by setting

\begin{equation}
\alpha_N = \alpha_1, 
\end{equation}

\noindent where $\alpha_1$ is the charge parameter of $\cO^{\, p, \, \pp}_{\i} (z_{\i})$, as 
in equation (\ref{minimal.model.vertex.operator}), which satisfies the FLW condition as in equation 
(\ref{charge.01}), and 
$\alpha_N$ is the charge parameter of $\cO^{\cH         }      (z_{\i})$, as
in equation (\ref{carlsson.okounkov}).

A holomorphic linear conformal block that consists of $n$ consecutive channels is the expectation 
value of $(n+3)$ holomorphic vertex operators at positions $z_i, i = 0, \cdots, n+3$, 

\be
\label{expectation.value}
\cB^{\, p, \, \pp}_{N, n} = 
\langle 
\, \cO_{\, \vec r_0      \, \vec s_0    }  (z_0    )
\, \cO_{\,      r_1       \,     s_1    }  (z_1    ) \cdots 
\, \cO_{\,      r_{n+1}   \,     s_{n+1}}  (z_{n+1})
\, \cO_{\, \vec r_{n+2}  \, \vec s_{n+2}}  (z_{n+2})
\, \rangle,
\ee
\noindent where the vertex operators $\cO_{\, r_i, \, s_i}$, $i=1, \cdots, n+1$, are specified below. 
When the positions $z_i$ are generic, global conformal invariance on the sphere can be used to set 
$z_0 = 0$, 
$z_{n+1} = 1$, $z_{n+2} = \infty$, then to scale the positions of the remaining points such that
\footnote{\,
We take $z_1$ to be closest to $z_0 = 0$, followed by $z_2$, {\it etc.} and $z_{n+2}$ to be farthest.
}
\begin{equation}
\label{ratio.positions}
q_\i = \frac{ | z_\i |}{ | z_{\i+1} | } < 1, \quad \i = 1, \cdots, n,
\end{equation}

\subsubsection{The parameters that appear in $\cB^{\, p, \, \pp}_{N, n}$} 
We can summarise the above discussion as follows.
$\cB^{\, p, \, \pp}_{N, n}$ depends on three sets of parameters
\footnote{\, 
As pointed out earlier, the subscript $i$ for $\vec r_i$ and $\vec s_i$ is the position of the corresponding 
operator insertion, and should not be confused with a component of the vector $r_i$ and $s_i$.
}, 

\begin{equation}
\label{conformal.block}
\cB^{\, p, \, \pp}_{N, n} =
\cB^{\, p, \, \pp}_{N, n} 
\ll  q_1,                             \cdots,   q_n \, | 
\, \vP_{\, \vec r_0     \, \vec s_0}, \cdots, \vP_{\, \vec r_{n+2} \, \vec s_{n+2}} \, | 
\, \va_{\,      r_1     \,      s_1}, \cdots, \va_{        r_{n+1} \,      s_{n+1}}
\rr
\end{equation}

\noindent The parameters $\{q_1, \cdots, q_n \}$ are the ratios of consecutive positions 
defined in (\ref{ratio.positions}).
The charge vectors $\vP_{\, \vec r_0 \, \vec s_0}$ and $\vP_{\, \vec r_{n+2} \, \vec s_{n+2}}$
label the $\cW_N$ initial and final states. They do not need to satisfy the FLW condition.
The charge vectors $\vP_{\, \vec r_\i \, \vec s_\i}$, $\i = 1, \cdots, n$, label 
the highest weight states of the $\cW_N$ irreducible highest weight representations that flow 
in the $\i$-th channel.
The charges $\va_{r_1 \, s_1}, \cdots, \va_{r_{n+1} \, s_{n+1}}$ label the vertex operator 
insertions. They need to satisfy the FLW condition, so as vectors on the $A_2$ weight lattice,
they point in th edirection of the fundamental weight $\vomega_1$ only.

\subsubsection{The Heisenberg factor} The conformal block $\cB^{\, p, \, \pp, \, \cH}_{N, n}$, 
which includes the contribution of the Heisenberg algebra, depends on the same parameters as 
$\cB^{\, p, \, \pp}_{N, n}$, in (\ref{conformal.block}). The two expressions are related by

\be
\label{conformal.block.refined}
\cB^{\, p, \, \pp, \, \cH}_{N, n} =
\prod_{\i= 1}^n 
\prod_{ l=\i}^n 
\ll 1 - q_{\i} \cdots  q_l \rr^{
\frac{
a_{\i+1} \ll \alpha_0 - a_{l+2} \rr
}
{
N
}
}
\,\,
\cB^{\, p, \, \pp}_{N, n},
\ee

\noindent where the variables $q_\i$, $\i = 1, \cdots, n$, were defined in (\ref{ratio.positions}),
and $a_{\i}$ is an abbreviation of $a_{r_{\i} s_{\i}}$. The factor that multiplies 
$\cB^{\, p, \, \pp}_{N, n}$, on the right hand side of (\ref{conformal.block.refined}), 
is the Heisenberg factor. It follows directly from the contribution of the Heisenberg vertex operators 
in (\ref{carlsson.okounkov}) to the expectation value in (\ref{expectation.value}).

%SECTION.03
\section{$\cW_N$ AGT correspondence}
\label{wn.agt}

{\it We recall basic definitions related to partitions, then discuss the $\cW_N$ AGT correspondence.}

\subsection{Partitions}
A partition $\pi$ of an integer $| \pi |$ is a set of non-negative integers $\{ \pi_1, \cdots,$ $\pi_p \}$, 
where $p$ is the number of parts, $\pi_i \geq \pi_{i+1}$, and $\sum_{i=1}^p \pi_i= |\pi|$. 
$\pi$ is represented by a Young diagram $Y$, which is a set of $p$ rows $\{Y_1, \cdots, Y_p\}$, such 
that the $\rrho$-th row has $Y_\rrho = \pi_\rrho$ cells, $Y_\rrho \geq Y_{\rrho + 1}$, and 
$| Y | = \sum_\rrho Y_\rrho = |\pi|$. We use $Y_\rrho$ for the $\rrho$-th row as well as for the number 
of cells in that row. $Y^{\intercal}$ is the transpose of $Y$. 

\subsubsection{Cells and coordinates} 
\label{cells.and.coordinates}
We use $\square$ for a cell in a Young diagram $Y$, which is a square in the south-east 
quadrant of the plane, with coordinates $\{\rrho, \ssigma\}$, such that 
$\rrho$   is the row-number,    counted from top  to bottom, and 
$\ssigma$ is the column number, counted from left to right. 

\subsubsection{Arms and legs}
\label{arms.and.legs}
$A_{\square, Y_i}$ is the arm of $\square$ in $Y_i$, 
that is, the number of cells in the same row as, but to the right of $\square$ in $Y_i$,
and
$L_{\square, W_j}$ to be the leg of $\square$ with respect its position in $W_j$, 
that is the number of cells in the same column as, but below $\square$ in $Y_i$. 
We define $A^+_{\square, Y_i} = A_{\square, Y_i} + 1$. 

%FIG.02
%
\begin{center}
\begin{minipage}{5.0in}
\setlength{\unitlength}{0.001cm}
\renewcommand{\dashlinestretch}{30}
\begin{picture}(4800, 3200)(-4000, 2000)
%
% negative x shifts the figure to the right
% positive x shifts the figure to the left
% negative y shifts the figure up
% positive y shifts the figure down
\thicklines
%
%The horizontal lines
%\path(0000,0000)(0600,0000)
%\path(0000,0600)(1200,0600)
%\path(0000,1200)(1200,1200)
 \path(0000,3600)(3000,3600)
 \path(0000,3000)(3000,3000)
 \path(0000,2400)(2400,2400)
%\path(0000,1800)(2400,1800)
%
\put(-0500,3150){$Y_1$}
\put(-0500,2550){$Y_2$}
%\put(-0500,1950){$Y_3$}
%\put(-0500,1350){$Y_4$}
%\put(-0500,0750){$Y_5$}
%\put(-0500,0150){$Y_6$}
%
%The vertical lines
\path(0000,3600)(0000,2400)
\path(0600,3600)(0600,2400)
\path(1200,3600)(1200,2400)
\path(1800,3600)(1800,2400)
\path(2400,3600)(2400,2400)
\path(3000,3600)(3000,3000)
\put(0050,3800){$Y^{\intercal}_1$}
\put(0650,3800){$Y^{\intercal}_2$}
\put(1250,3800){$Y^{\intercal}_3$}
\put(1850,3800){$Y^{\intercal}_4$}
\put(2450,3800){$Y^{\intercal}_5$}
\put(0750,2600){$\checkmark$}
\end{picture}
\begin{ca}
\label{1.Young.diagram}
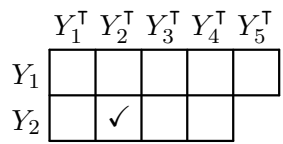
The 2-row           Young diagram $Y$             of the partition $5 + 4$.
The rows are numbered from top to bottom. 
The 5-row transpose Young diagram $Y^{\intercal}$ of the partition $2 + 2 + 2 + 2 + 1$, 
which are the columns of $Y$, are numbered from left to right. 
From the viewpoint of $Y$, the marked cell $\checkedcell$ has 
$A^{ }_{\checkedcell, Y} = 2$, 
$A^{+}_{\checkedcell, Y} = 3$, 
and 
$L_{\checkedcell, Y} = 0$.
From the viewpoint of $Y^{\intercal}$,     $\checkedcell$ has
$A^{ }_{\checkedcell, Y^{\intercal}} = 0$, 
$A^{+}_{\checkedcell, Y^{\intercal}} = 1$, 
and
$L_{\checkedcell, Y^{\intercal}} = 2$.
\end{ca}
\end{minipage}
\end{center}
\bigskip

\subsubsection{$N$-partitions}
The AGT representation of $\cB^{\, p, \, \pp, \, \cH}_{N, n}$ involves a multi-sum over $(n+2)$ 
$N$-partitions 
$\vec Y^{\i}$, $\i = 0, \cdots, n+1$, where $\vec Y^{\i}$ is a set of of $N$ Young diagrams, 
$\{ Y_1^{\i}, \cdots, Y_N^{\i} \}$, and
$|\vec Y^{\i }| = |Y_1^{\i}| + \cdots + |Y_N^{\i}|$ is the total number of cells in 
$\vec Y^{\i}$.
The $N$-partitions $\{ Y^{\i }_1, \cdots, Y^{\i }_N \}$, $\i \in 1, \cdots, n$, are 
non-empty Young diagrams, while $\{ Y^{\i }_1, \cdots,$ $Y^{\i }_N \}$, $\i = 0, n+1$ are empty 
\footnote{\, 
We work in terms of $(n+2)$ linearly-ordered $N$-partitions. Since we consider conformal blocks 
of vertex operators, the initial and final $N$-partitions are always empty, but we prefer to work 
in terms of $(n+2)$ rather than $n$ non-empty $N$-partitions to make the notation in the sequel 
more uniform.
}, 
$\vec Y^{(0)} = \vec Y^{(n+1)} = {\vec \emptyset}$, where $\vec \emptyset$ is an $N$-partition that 
consists of $N$ empty Young diagrams. 

%FIG.03
%
\begin{center}
\begin{minipage}{5.0in}
\setlength{\unitlength}{0.001cm}
\renewcommand{\dashlinestretch}{30}
\begin{picture}(4800, 2600)(-2000, 2000)
% negative x shifts the figure to the right
% negative y shifts the figure up
% positive y shifts the figure down
\thicklines
%
%Horizontal to the LEFT
 \path(0000,3600)(3000,3600)
 \path(0000,3000)(3000,3000)
 \path(0000,2400)(2400,2400)
%\path(0000,1800)(2400,1800)
%\path(0000,1200)(1200,1200)
%\path(0000,0600)(1200,0600)
%\path(0000,0000)(0600,0000)
%
%Vertical on the LEFT
 \path(0000,3600)(0000,2400)
 \path(0600,3600)(0600,2400)
 \path(1200,3600)(1200,2400)
 \path(1800,3600)(1800,2400)
 \path(2400,3600)(2400,2400)
 \path(3000,3600)(3000,3000)
%
%Horizontal to the RIGHT
 \path(3900,3600)(6300,3600)
 \path(3900,3000)(6300,3000)
%\path(3900,2400)(4500,2400)
%\path(3900,1800)(4500,1800)
%\path(3900,1200)(4500,1200)
%\path(3900,0600)(4500,0600)
%
%Vertical to the RIGHT
 \path(3900,3600)(3900,3000)
 \path(4500,3600)(4500,3000)
 \path(5100,3600)(5100,3000)
 \path(5700,3600)(5700,3000)
 \path(6300,3600)(6300,3000)
\put(1400,2600){$\checkmark$}
\put(5300,2600){$\checkmark$}
\end{picture}
\begin{ca}
\label{2.Young.diagrams}
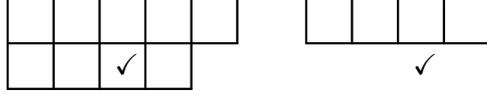
A 2-partition $\{Y_1, Y_2\}$. $Y_1$ is on the left, $Y_2$ is on the right. The cell $\checkedcell$
has coordinates $(2, 3)$,  
$A^{ }_{\checkedcell, Y_1} =  1$,
$A^{+}_{\checkedcell, Y_1} =  2$,
$L^{ }_{\checkedcell, Y_1} =  0$,
$A^{ }_{\checkedcell, Y_2} = -3$,
$A^{+}_{\checkedcell, Y_2} = -2$,
and
$L^{ }_{\checkedcell, Y^1} = -1$.
\end{ca}
\end{minipage}
\end{center}
\bigskip

\subsection{Extending AGT to $\cW_N$}
\label{extending.wn.agt}

The AGT correspondence of Alday, Gaiotto and Tachi\-kawa \cite{agt}, extended to $\cW_N \oplus 
\cH$ by Mironov and Morozov \cite{mironov.morozov} and Wyllard \cite{wyllard}, identifies a class
of conformal blocks in $\cM_N^{\, non.min, \cH}$, that we specify below, with instanton 
partition functions in 4-dimensional $\cN \! = \! 2$ supersymmetric quiver $U(N)$ gauge theories 
\cite{nekrasov}. 

\subsubsection{AGT in non-minimal $\cW_N$ models} 
The $\cW_N$ AGT expression for a conformal block $\cB^{\, non.min, \cH}_{N, n}$, 
that has $n$ consecutive channels $\chi_{\i}$, $\i = 1, \cdots, n$, is an $n$-fold sum
\footnote{\, 
Recall that the $N$-partitions $\vec Y^0$ and $\vec Y^{n+1}$ are empty.
}, 

\begin{equation}
\label{agt.gen}
\cB^{\, non.min, \cH}_{N, n}=\sum_{\vec Y^1, \cdots, \vec Y^n}
\prod_{\i = 1}^{n+1} q_{\i}^{|  \vec Y^{\i    }  |} 
Z_{bb}^{\i} \ll \vP_{(\i - 1)}, \vec Y^{\i - 1}\ | \ a_{\i } \ | \ \vP_{(\i)}, \vec Y^{\i} \rr,
\end{equation}

\noindent the factors 
$q_{\i}^{| \vec Y^{\i} |} Z_{bb}^{\i} [ \vP^{(\i - 1)}, \vec Y^{\i - 1}\ | \ \mu^{\i} \ | \ \vP^{(\i)}, 
\vec Y^{\i}]$, $\i = 1, \cdots, n+1$, are defined in subsection {\bf \ref{agt.non.minimal}}. Each 
factor $Z_{bb}^{\i}$ is a rational function that depends on two $N$-partitions of {\it \lq unrestricted\rq} 
Young diagrams $\{Y^{\i - 1}_1, \cdots, Y^{\i - 1}_N \}$ and $\{Y^{\i}_1, \cdots, Y^{\i}_N \}$. In 
other words, there are no conditions on these Young diagrams and all possible $N$-partitions are allowed. 
The denominator $z_{den}^{\i}$ of $Z_{bb}^{\i}$ is a product of the norms of the states that flow in the 
preceding channel $\chi^{\i - 1}$ and the subsequent channel $\chi^{\i}$.
Since $Z_{bb}^{\i}$ is labeled by unrestricted $N$-partitions, and the sums are over all possible unrestricted 
$N$-partitions, the states that flow in each channel belong to a Verma module of $\cW_N^{\, non.min, \cH}$.

\subsubsection{AGT for non-minimal $\cW_N$ }
\label{agt.non.minimal}

The decomposition of conformal blocks in (\ref{agt.gen}) follows that in \cite{kanno.matsuo.zhang} and is 
represented as a {\it comb diagram} in Figure {\bf \ref{the.comb.diagram}}. The function $Z_{bb}$ is 

\begin{equation}
\label{z.bb}
Z_{bb}  \ll \vec a, \vec Y\ | \ \mu \ | \ \vec b, \vec W \rr = 
\frac{
z_{num} \ll \vec a, \vec Y\ | \ \mu \ | \ \vec b, \vec W \rr
}
{
z_{den} \ll \vec a, \vec Y\ |           \ \vec b, \vec W \rr
},
\end{equation}

\noindent and has the following ingredients. 
$N$-component vector 
${\vec a^{\i}} = \{ a_1^{\i}, \cdots, a_N^{\i} \}$, such that 
$\sum_{i=1}^N $ $a_i^{\i} = 0$,  is the charge of the highest weight state of 
the $\cW_N$ irrep that flows in the intermediate channel $\chi_{\i}$.
Each of the two $N$-partition sets 
$\vec V^{\i}$ $= \{ V_1^{\i}, \cdots$, $V_N^{\i}\}$, and 
$\vec W^{\i}$ $= \{ W_1^{\i}, \cdots$, $W_N^{\i}\}$, 
labels the elements of the special orthogonal basis in the $\cM_N \! \times \! \cM^{\cH}$ 
Verma module associated with the vertex operator in channel $\chi_{\i}$. 
In Equation {\bf \ref{agt.gen}}, $\vec Y$ and $\vec W$ are attached to the line segments 
on the left and the right of a given vertex, respectively, see Figure {\bf \ref{the.comb.diagram}}.
The scalar $\mu^{\i}$ is the charge of the vertex operator that connects channels 
$\chi_{\i}$ and $\chi_{\i + 1}$. 
In the following, we study the structure of the right hand side of (\ref{z.bb}).

\subsubsection{The numerator} 
\label{the.numerator}
\begin{multline}
\label{z.num}
z_{num} \ll \vec a, \vec Y\  | \ \mu \ | \ \vec b, \vec W \rr = 
\\
\prod_{i, j = 1}^N 
\prod_{\square \in Y_i}
\ll
E[a_i - b_j, Y_i, W_j, \square] - \mu
\rr
\prod_{\blacksquare \in W_j}
\ll
\epsilon_1 + \epsilon_2 - E[b_j - a_i, W_j, Y_i, \blacksquare] - \mu
\rr,
\end{multline}
\noindent where the elementary function $E[x, Y_i, W_j, \square]$ is defined as

\begin{equation}
\label{E.non.minimal.1}
E[x_{ij}, Y_i, W_j, \square] = 
x_{ij} 
+ A^{+}_{\square, Y_i} \epsilon_2 
- L^{ }_{\square, W_j} \epsilon_1, 
\end{equation}
\noindent $x_{ij}$ is an indeterminate, and $\{\epsilon_1, \epsilon_2\}$ are complex 
parameters related to the central charge to be specified below.

\subsubsection{The denominator} 
\label{the.denominator}
\begin{equation}
\label{z.den}
z_{den} \ll \vec a, \vec Y \ | \ \vec b, \vec W \rr
= 
\ll 
z_{norm} \ll \vec a, \vec Y \rr \ 
z_{norm} \ll \vec b, \vec W \rr 
\rr^{\frac{1}{2}},
\end{equation}

\begin{equation}
\label{z.norm}
z_{norm} \ll \vec a, \vec Y \rr = z_{num} \ll \vec a, \vec Y \ | \  0 \ | \ \vec a, \vec Y \rr 
\end{equation}

\noindent In gauge theory, $z_{norm}$ is a normalization factor related to the contribution of 
the vector multiplets \cite{agt}. 
In conformal field theory, it accounts for the norms of the states that propagate into and out 
of the vertex operator insertion in $Z_{bb}$. 

%SECTION.04
\section{AGT for minimal models. Finiteness and the $N$-Burge conditions}
\label{agt.minimal}

{\it We consider the building block partition function introduced in (\ref{z.bb}) and 
subsequent equations, and set the parameters to those relevant to $\cM_N^{\, p, \, p', \, \cH}$.
We show that by restricting the Young diagrams, we obtain well-defined expressions that we 
identify with minimal $\cW_N$ conformal blocks.}

\subsection{Minimal model parameters}
Since we focus on the minimal models, we choose to work in terms of the screening charge parameters 
$\{ \aneg, \apos \}$ rather than Nekrasov's deformation parameters $\{ \epsilon_1, \epsilon_2 \}$, 
by setting 

\begin{equation}
\label{neg.0.pos}
\aneg = \epsilon_1, 
\quad
\apos = \epsilon_2, 
\end{equation}

\noindent where $\aneg$ and $\apos$ are real and satisfy $\aneg < 0 < \apos$. We write 

\begin{equation}
\label{E.minimal}
\boxed{
E[x_i, x_j, Y_i, W_j, \square] =
x_i - x_j
+ A^{+}_{\square, i}  \, \apos
- L^{ }_{\square, j}  \, \aneg
}
\end{equation}

\subsection{$\cW_N$ parameters}
The parameters $x_i$ and $x_j$ in (\ref{E.minimal}) are scalar components of the vector of gauge 
theory Coulomb parameters $\{x_1, \cdots, x_N\}$, that satisfy $\sum_{i=1}^N x_i = 0$ and 
$A^+_{\square, i} = A_{\square, i} + 1$. We identify the Coulomb parameters with the minimal 
model parameters by setting 

\begin{equation}
x_i = x^+_i \apos
    + x^-_i \aneg, \quad i = 1, \cdots, N
\end{equation}

\noindent and choosing 

\begin{equation}
x^{+}_i =  - \sum_{j=1}^{N-1}  \langle \vomega_j \,| \, \vh_i\rangle \, r_j, \quad
x^{-}_i =  - \sum_{j=1}^{N-1}  \langle \vomega_j \,| \, \vh_i\rangle \, s_j
\end{equation}

\noindent where $\vomega_i$, $i = 1, \cdots, N-1$, are the $A_{N-1}$ fundamental weight vectors, 
$\vh_i$, $i = 1, \cdots, N-1$, are the weight vectors of the first fundamental representation 
of the Lie algebra $A_{N-1}$, and 
$\langle \vomega_j \, | \, \vh_i \rangle$ is the scalar product of $\vomega_j$ 
and $\vh_i$, regarded as $N$-component vectors in the weight lattice of $A_{N-1}$. 
Noting that $\vh_i - \vh_{i+1} = \valpha_i$, $i = 1, \cdots, N-1$, where $\valpha_i$ are the simple 
root vectors of $A_{N-1}$, and that $\langle \vomega_i | \valpha_j \rangle = \delta_{ij}$, where 
$\delta_{ii} = 1$, and $\delta_{ij} = 0$, for $i \neq j$, the above definitions allow us to write 

\begin{equation}
\label{from.x.to.r.and.s}
\boxed{
x^{+}_i - x^{+}_{i+1} = - r_i, \quad
x^{-}_i - x^{-}_{i+1} = - s_i, \quad
i = 1, \cdots, N-1
}
\end{equation}

\subsection{Scanning products for zeros}
Consider the denominator $z_{den}$ of $Z_{bb}$, defined in {\bf \ref{the.denominator}}. To look 
for zeros in $z_{den}$, it is sufficient to look for zeros in $z_{norm} [\vec x, \vec Y]$, defined 
in (\ref{z.norm}). Consider $\cB^{\, p, \, \pp, \, \cH}_{N, n}$ and focus on a channel that carries 
states that belong to the degenerate $\cW_n$ irreducible highest weight representation 
$\cH^{\, p, \, \pp}_{r, s}$, where $p$ and 
$\pp$ are coprime, $0 < p < \pp$, $r = r_1, \cdots, r_{N-1}$, and $s = s_1, \cdots, s_{N-1}$. Recall 
that we also define $r_N = p - \sum_{i=1}^{N-1} r_i$, and $s_N = \pp - \sum_{i=1}^{N-1} s_i$, and that 
$0 < r_i < p$, and $0 < s_i < \pp$, $i = 1, \cdots, N$.

\begin{Proposition} 
\label{proposition.01}
$z_{norm} [\vec x, \vec Y] \neq 0$, if and only if 

\begin{equation}
\label{conditions.01.simpler}
\boxed{
Y_{i+1,  \, \rrho} - Y_{i,  \, \rrho +  s_i - 1} \geq - r_i + 1
}
\end{equation}

\noindent where $Y_{i,  \, \rrho}$ is the $\rrho$-row in $Y_i$, $i = 1, \cdots, N$, $r_i$ and $s_i$, 
$i = 1, \cdots, N$, are the integers that parameterise the degenerate $\cW_N$ irreducible highest 
weight representation that flows in the channel under consideration,
$r_N =   p - \sum_{i=1}^{N-1} r_i$ and 
$s_N = \pp - \sum_{i=1}^{N-1} s_i$. 
\end{Proposition}

\subsection{Zero-conditions}
The proof of Proposition {\bf \ref{proposition.01}} is based on checking the factors that appear in 
$z_{norm} [\vec x, \vec Y]$ for zeros. This requires introducing a number of elementary concepts 
that were first introduced in \cite{bershtein.foda}.

\subsubsection{Two zero-conditions}
\label{to.vanish}
As we will show, a factor in $z_{norm}$ has a zero when an equation of type

\begin{equation}
\label{zeros.example.01}
C_{+} \, \apos + C_{-} \, \aneg = 0,
\end{equation}

\noindent is satisfied, where $C_+$ and $C_-$ are non-zero positive integers, and $\aneg < 0 < \apos$. 
Since 
$\aneg = - p / \sqrt{p \, \pp}$, 
$\apos = \pp / \sqrt{p \, \pp}$, 
$p$ and $\pp$ are coprime, the same factor in $z_{norm}$ has a zero when the two conditions

\begin{equation}
\label{zeros.example.02}
C_+ =  c \   p, 
\quad
C_- =  c \ \pp, 
\end{equation}

\noindent are satisfied, where $c$ is a proportionality constant that remains to be determined. 

\subsubsection{From two zero-conditions to one zero-condition} 
\label{from.2.to.1}
This paragraph contains the core of the proof. Consider the two conditions

\begin{equation}
\label{conditions.example.01}
\boxed{
  A^{+}_{\square, i} =  A^{\prime} \geq 1,
\quad 
- L^{ }_{\square, j} =  L^{\prime} \geq 1
}
\end{equation}

\noindent where $A^+_{\square, i} = A_{\square, i} + 1$. These conditions are satisfied 
{\it if and only if} $\square \in Y_i$, and $\square \not \in Y_j$.
If $\square$ is in row-$\rrho$ and column-$\ssigma$ in $Y_i$, then the first condition in 
(\ref{conditions.example.01}) implies that there is a cell $\boxplus \in Y_i$, that may be
$\square$ or lies to the right of $\square$, with coordinates $\{\rrho, \ssigma + A^{\prime} - 1\}$, 
such that, this cell $\boxplus$ lies on a vertical boundary of $Y_i$. 
The latter statement means that, {\bf 1.} there are no cells to the right of $\boxplus$, and 
{\bf 2.} there may or may not be cells below $\boxplus$. 
The latter two statements imply that 
the $[\ssigma + A^{\prime} - 1]$-column in $Y_i^{         }$, or equivalently, 
the $[\ssigma + A^{\prime} - 1]$-row    in $Y_i^{\intercal}$, 
has length {\it at least} $\rrho$, 

\begin{equation}
\label{conditions.example.02}
Y^{\intercal}_{i, \, \ssigma + A^{\prime} - 1} \geq \rrho
\end{equation}

\noindent Using the definition of $L_{\square, j}$ in {\bf \ref{arms.and.legs}}, we see that 
$L_{\square, j} = Y^{\intercal}_{j, \, \ssigma} - \rrho$, and we can write the equality in 
the second condition in (\ref{conditions.example.01}) as 

\begin{equation}
\label{conditions.example.03}
- L_{\square, j} = - Y^{\intercal}_{j, \, \ssigma} + \rrho = L^{\prime} 
\end{equation}

\noindent In other words,  

\begin{equation}
\rrho = L^{\prime} + Y^{\intercal}_{j, \, \ssigma},  
\end{equation}

\noindent and from (\ref{conditions.example.02}), we obtain  

\begin{equation}
\label{conditions.example.04}
Y^{\intercal}_{i, \, \ssigma + A^{\prime} - 1} - Y^{\intercal}_{j, \, \ssigma} \geq L^{\prime},
\end{equation}

\noindent where $L^{\prime} > 0$, which is one condition that is equivalent to the two conditions 
in (\ref{conditions.example.01}). 

\subsection{Non-zero condition}
\label{1.not.to.vanish}
Consider a function $\cF [Y_i, Y_j]$, of a pair of Young diagrams $\{Y_i, Y_j\}$, such that 
$\cF [Y_i, Y_j] = 0$, {\it if and only if} (\ref{conditions.example.04}) is satisfied. This 
implies that $\cF [Y_i, Y_j] \neq 0$, {\it if and only if} $\{Y_i, Y_j\}$ satisfies the 
complementary condition

\begin{equation}
\label{conditions.example.05}
Y^{\intercal}_{i, \, \ssigma + A^{\prime} - 1} - Y^{\intercal}_{j, \, \ssigma} < L^{\prime},
\end{equation}

\noindent which we choose to write as 

\begin{equation}
\label{conditions.example.06}
Y^{\intercal}_{j, \, \ssigma} - Y^{\intercal}_{i, \, \ssigma + A^{\prime} - 1} \geq 1 - L^{\prime} 
\end{equation}

\noindent which, following \cite{bershtein.foda}, can be written in the form
\footnote{\, 
The proof that (\ref{conditions.example.06}) is equivalent to (\ref{conditions.example.07}) is in 
subsection {\bf 4.10} of \cite{bershtein.foda}.
}

\begin{equation}
\label{conditions.example.07}
\boxed{
Y_{j,  \, \rrho} - Y_{i,  \, \rrho + L^{\prime} - 1} \geq 1 - A^{\prime}
}
\end{equation}

\subsubsection{Remarks} 
{\bf 1.} It is useful, for the purposes of the calculations in the sequel, to note that re-writing 
(\ref{conditions.example.06}) as (\ref{conditions.example.07}) is equivalent to transposing each 
of the partitions $Y_i$ and $Y_j$, replacing the shift in the row number of $Y_j$ by the negative 
of the right hand side, and {\it vice versa}.
{\bf 2.} The subscripts $\ssigma$ and $\ssigma + A^{\prime} - 1$ on the left hand side of 
(\ref{conditions.example.06}) refer to {\it the row-numbers} of the Young diagrams $Y^{\intercal}_j$ 
and $Y^{\intercal}_i$, respectively.
{\bf 3.} The Young diagram that the cell $\square$ lives in, in this case $Y_i$, appears with a minus 
sign in (\ref{conditions.example.07}). We frequently meet such equations in the sequel, and we need 
such observations to be able to make simple, quick checks of their consistency.
{\bf 4.} We refer, in the sequel, to equations such as (\ref{conditions.example.01}) and 
(\ref{conditions.example.06}) as {\it \lq zero-conditions\rq}, and {\it \lq non-zero-conditions\rq}, 
respectively. 

\subsection{Products in the denominator} 
\label{products.denominator}
Two types of products appear in $z_{norm}$. These are
{\bf 1.} products in the form 
$\prod_{\square      \in Y_i}                  E[a_i - a_j, Y_i, Y_j, \square]$, 
that we denote by $\{Y_i, Y_j\}$, and 
{\bf 2.} products in the form 
$\prod_{\square      \in Y_i} [\apos + \aneg - E[a_i - a_j, Y_i, Y_j, \square]]$
that we denote by $\{Y_i, Y_j\}^{\prime}$. 

\subsubsection{Remark} As we will show, it is sufficient to consider $\{Y_i, Y_{i+1}\}$, 
$i = 1, \cdots, N$, where $Y_{N+1} = Y_1$. The conditions that remove these zeros are sufficient to 
remove the zeros of the other products.

\subsubsection{In search of zeros}
\label{in.search.of.zeros}
We plan to proceed as follows. 
{\bf 1.} We consider the products in $z_{den}$, one at a time, 
{\bf 2.} search for possible zeros, as in subsection {\bf \ref{to.vanish}}, 
{\bf 3.} find the conditions that imposed on the pair $\{Y_i, Y_j\}$ in order to avoid the zeros, and 
{\bf 4.} when there is more than one of condition to avoid a zero, we choose the strongest condition. 
That is, we choose the condition that eliminates more zeros than any other condition. To do this, we 
use the fact that $r_i$, and $s_i$, $i = 1, \cdots, N$, are non-zero positive integers.

\subsubsection{Products that have no zeros}
\label{products.that.have.no.zero}
$\{Y_i, Y_i\}$, $i = 1, \cdots, N$, has no zeros since that requires a factor that satisfies

\begin{equation}
\label{zeros.11}
E[0, Y_i, Y_i, \square] 
=
  A^{+}_{\square, i} \  \apos
- L^{ }_{\square, i} \  \aneg
=
0,
\end{equation}

\noindent which is not possible since $\square \in Y_i$, thus 
$A^{+}_{\square, i}    > 0$,  
$L^{ }_{\square, i} \geq 0$,
$\apos > 0$, and 
$\aneg < 0$. 
$\{Y_i, Y_i\}^{\prime}$, $i = 1, \cdots, N$, has no zeros for the same reason. 

\subsubsection{Products that have zeros}
Next, we consider $\{Y_i, Y_j\}$, such that $i \neq j$, of which there are $N (N-1)$ products. 
To do that, we need to introduce some simple definitions.

\subsubsection{Periodic $N$-partitions} 
\label{periodic.N.partitions}
It is convenient to regard the set of $N$-partitions $\{ Y_1, \cdots$, $Y_N \}$ as a subset 
of a set of infinitely-many partitions with periodicity $N$. More precisely, we consider 
a set of infinitely-many partitions, $Y_i$, $i \in \ZZ$, and define $Y_{i+kN} = Y_i$, 
$i \in 1, \cdots, N$, $k=\ZZ$. 
The $N$-partitions that we start with correspond to the {\it \lq fundamental subset\rq} $Y_i$, 
$i = 1, \cdots, N$
\footnote{\,
As explained in paragraph {\bf {\ref{cylindric.N.partitions}}}, we actually end up with 
cylindric partitions \cite{gessel.krattenthaler}, since adjacent partitions $Y_i$ and $Y_{i+1}$, 
$i \in \ZZ$, are related by conditions of the type discussed in \cite{gessel.krattenthaler}. 
}. 

\subsubsection{An $N$-site circle $\cC_N$} 
\label{N.circle}
Since $Y_i = Y_{i+kN}$, $i \in 1, \cdots, N$, $k \in \ZZ$, we regard $Y_i$, $i \in \ZZ$ as 
assigned to the sites $\sigma_i$, $i=1, \cdots, N$, of an $N$-site circle $\cC_N$, and assign 
partition $Y_{i+ k N}$ to site $i\sigma_i$.

\subsubsection{Periodic $x_i^+$, $x_i^-$, $r_i$ and $s_i$ parameters}
Similarly to the partitions $Y_i$, $i \in 1, \cdots, N$, whose definition is extended to all 
$i \in \ZZ$, we extend the definition of the parameters $x_i^+$, $x_i^-$, $r_i$ and $s_i$, 
$i, 1, \cdots, N$, defined in {\bf \ref{primary.fields}}, to $x_i^+$, $x_i^-$, $r_i$ and $s_i$, 
$i \in \ZZ$, and define 

\begin{equation}
x^+_{i+kN} = x^+_i,
\quad
x^-_{i+kN} = x^-_i,
\quad
r_{i+kN} = r_i, 
\quad
s_{i+kN} = s_i, 
\quad
i = 1, \cdots, N, 
\quad
k \in \ZZ
\end{equation}

\noindent We attach $x^+_{i+kN}$, $x^-_{i+kN}$, $r_{i+kN}$ and $s_{i+kN}$, $i = 1, \cdots, N$, 
$k \in \ZZ$, to site $\sigma_i$ in $\cC_N$. 
The parameters that we start with correspond to those in the {\it fundamental subset} $x^+_i$, 
$x^-_i$, $r_i$ and $s_i$, $i=1, \cdots, N$. 

\subsection{Conditions from $\{Y_i, Y_{i+1} \}$}
\label{i.i+1}
Each of these products, for $i = 1, \cdots, N$, vanishes if it contains a factor that satisfies 

\begin{multline}
\label{i.i+1.01}
E[ x_i, x_{i+1}, Y_i, Y_{i+1}, \square]
=
\\
\ll x^+_i - x^+_{i+1} + A^{+}_{\square, i  } \rr \, \apos
+
\ll x^-_i - x^-_{i+1} - L^{ }_{\square, i+1} \rr \, \aneg
=
\\
\ll - r_i + A^{+}_{\square, i  } \rr \, \apos
+
\ll - s_i - L^{ }_{\square, i+1} \rr \, \aneg
=
0,
\end{multline}
\noindent which, using (\ref{from.x.to.r.and.s}), leads to the conditions
\begin{equation}
\label{i.i+1.02}
  A^{+}_{\square, i  } = r_i + c\,   p, 
\quad 
- L^{ }_{\square, i+1} = s_i + c\, \pp 
\end{equation}

\noindent where $c$ remains to be determined. Since $A_{\square, i}$, $L^{ }_{\square, i+1}$,
$r_i$ and $s_i$, $i = 1, \cdots, N$, are non-zero positive integers, and $p$ and $\pp$ are 
positive co-primes, $c$ must be an integer. Since $r_i < p$, $i = 1, \cdots, N$, if $c < 0$, 
$A^{+}_{\square, i  } < 0$, which is not possible, hence $c = 0, 1, 2, \cdots$ 
In other words, conditions (\ref{i.i+1.02}) are possible for $c = 0, 1, 2, \cdots$, $\square 
\in Y_i$ and $\square \not \in Y_{i+1}$.
\subsubsection{From two zero-conditions to one non-zero-condition} 
\label{translating.01}
Following paragraph {\bf \ref{from.2.to.1}} and subsection {\bf \ref{1.not.to.vanish}}, the 
two zero-conditions in (\ref{i.i+1.02}) can be translated to one non-zero-condition, 
\begin{equation}
\label{i.i+1.03}
Y^{\intercal}_{i+1, \, \ssigma} - Y^{\intercal}_{i, \, \ssigma + \ll r_i - 1 \rr + c\, p} \geq 
- \ll s_i - 1 \rr - c\, \pp 
\end{equation}
\subsubsection{The strongest condition} 
\label{the.strongest.condition}
Since the row-lengths of a partition are weakly decreasing, condition (\ref{i.i+1.03}) is 
satisfied if 

\begin{equation}
\label{i.i+1.04}
Y^{\intercal}_{i+1, \, \ssigma} - Y^{\intercal}_{i, \, \ssigma + \ll r_i - 1 \rr}         \geq 
- \ll s_i - 1 \rr - c\, \pp
\end{equation}

\noindent which is the case if 

\begin{equation}
\label{i.i+1.05}
Y^{\intercal}_{i+1, \, \ssigma} - Y^{\intercal}_{1, \, \ssigma + \ll r_i - 1 \rr} \geq - \ll s_i - 1 \rr
\end{equation}

\noindent Thus, we should set $c=0$, which is an allowed value for $c$, in (\ref{i.i+1.03}), and 
following \cite{bershtein.foda}, re-write it in the simpler form 
\begin{equation}
\label{i.i+1.06}
\boxed{
Y_{i+1,  \, \rrho} - Y_{i,  \, \rrho + \ll s_i - 1 \rr} \geq - \ll r_i - 1 \rr
}
\end{equation}

\subsection{Conditions from 
$\{ Y_{i    }, Y_{i+1} \}$,
$\{ Y_{i  +1}, Y_{i+2} \}$,
$\cdots$,
$\{Y_{i+n-1}, Y_{i+n} \}$
}
\label{n.i.i+1}
Since condition (\ref{i.i+1.06}) relates the partitions $Y_i$ and $Y_{i+1}$, it also relates, by 
adjacency, the partitions $Y_i$ and $Y_{i+n}$, $n \in \ZZ > 1$. For example, the two conditions 

\begin{equation}
\label{n.i.i+1.01}
Y_{i+2,  \, \rrho}  -  Y_{i+1,  \, \rrho + s_{i+1} - 1} \geq - r_{i+1} - 1,
\quad
Y_{i+1,  \, \rrho}  -  Y_{i,    \, \rrho + s_{i  } - 1} \geq - r_{i  } - 1 
\end{equation}

\noindent imply 

\begin{equation}
\label{n.i.i+1.02}
Y_{i+2,  \, \rrho} - Y_{i,  \, \rrho + \ll \sum_{j=0}^1 s_{i+j} \rr - 2} \geq 2 - \sum_{j=0}^1 r_{i+j} 
\end{equation}

\noindent and, in the same way, the $n$ adjacent $\{Y_i, Y_{i+1} \}$ conditions imply

\begin{equation}
\label{n.i.i+1.03}
Y_{i+n,  \, \rrho} - Y_{i,  \, \rrho - n + \ll \sum_{j=0}^{n-1} s_{i+j} \rr} \geq 
                                       n - \ll \sum_{j=0}^{n-1} r_{i+j} \rr
\end{equation}

\noindent We refer to condition (\ref{n.i.i+1.03}) as an {\it \lq n-adjacent\rq} $\{Y_i, Y_{i+1}\}$ 
condition, since it comes from $n$ conditions of type $\{Y_i, Y_{i+1}\}$ that involve $(n+1)$ adjacent 
partitions.

\subsubsection{Remarks} 
{\bf 1.} Note the shift by $-n$ of the row-number of partition $Y_i$ on the left hand side of 
(\ref{n.i.i+1.03}), and by $n$ of the term on the right hand side. Condition (\ref{n.i.i+1.03}) 
makes sense since 
$- n + \sum_{j=0}^{n-1} s_{i+j} \geq 0$, and 
$  n - \sum_{j=0}^{n-1} r_{i+j} \leq 0$. 
{\bf 2.} In the following, we show that it is sufficient to impose condition (\ref{n.i.i+1.01}) to eliminate 
the zeros of $\{Y_i, Y_{i+n}\}$, rather than any condition obtained from any another product involving these 
two partitions. 
Since condition (\ref{n.i.i+1.01}) follows from the $\{Y_i, Y_{i+1} \}$ conditions (\ref{i.i+1.05}), the latter 
are sufficient to eliminate the zeros in $\{Y_i, Y_{i+n}\}$. 

\subsubsection{A consistency check} $\{Y_i, Y_{i+N}\}$ leads to the condition

\begin{equation}
\label{consistency.01}
Y_{i+N,  \, \rrho} - Y_{i,  \, \rrho + \sum_{j=0}^{N-1} \ll s_{i+j} - 1 \rr} \geq 
                                     - \sum_{j=0}^{N-1} \ll r_{i+j} - 1 \rr
\end{equation}

\noindent which can be written, using (\ref{sum.conditions}), as

\begin{equation}
\label{consistency.02}
Y_{i,  \, \rrho} - Y_{i,  \, \rrho + \ll \pp - N \rr} \geq - \ll p - N \rr
\end{equation}

\noindent which are trivial conditions on $Y_i$, $i = 1, \cdots, N$, since $\pp > p \geq N$, by definition 
of the $\cW_N$ minimal models. This agrees with the fact that such products do not have zeros, and therefore 
should not be restricted by any conditions.

\subsection{Conditions from $\{Y_i, Y_{i+n}\}$}
\label{i.i+n}
Each of these products, for $i = 1, \cdots, N$, and $n > 0$, vanishes if it contains a factor that satisfies

\begin{multline}
\label{i.i+n.01}
E[ x_i, x_{i+n}, Y_i, Y_{i+n}, \square]
=
\\
\ll x^+_i - x^+_{i+n} + A^{+}_{\square, i  } \rr \, \apos
+
\ll x^-_i - x^-_{i+n} - L^{ }_{\square, i+n} \rr \, \aneg
=
\\
\ll   A^{+}_{\square, i  } - \sum_{j=0}^{n-1} r_{i+j} \rr \, \apos
+
\ll - L^{ }_{\square, i+n} - \sum_{j=0}^{n-1} s_{i+j} \rr \, \aneg
=
0,
\end{multline}

\noindent which leads to the conditions
\begin{equation}
\label{i.i+n.02}
  A^{ }_{\square, i  } = c\   p - 1 + \sum_{j=0}^{n-1} r_{i+j}, 
\quad
- L^{ }_{\square, i+1} = c\ \pp     + \sum_{j=0}^{n-1} s_{i+j}, 
\end{equation}
\noindent where $c$ remains to be determined. Following the same arguments used in {\bf \ref{i.i+1}}, 
$c$ must be a non-negative integer. In other words, conditions (\ref{i.i+n.02}) are possible for 
$c = \{0, 1, \cdots\}$, $\square \in Y_i$ and $\square \not \in Y_{i+n}$.
\subsubsection{From two zero-conditions to one non-zero-condition}
\label{translating.02}
Following paragraphs {\bf \ref{from.2.to.1}} and {\bf \ref{1.not.to.vanish}}, 
the two zero-conditions in (\ref{i.i+n.02}) can be translated 
to one non-zero-condition,
\begin{equation}
\label{i.i+n.03}
Y^{\intercal}_{i+n, \, \ssigma} - Y^{\intercal}_{i, \, \ssigma + c\,   p - 1 + \sum_{j=0}^{n-1} r_{i+j}} 
\geq                                                   1 - c\, \pp     - \sum_{j=0}^{n-1} s_{i+j} 
\end{equation}
\subsubsection{The strongest condition}
Following the arguments in paragraph {\bf \ref{the.strongest.condition}}, the strongest version 
of condition (\ref{i.i+n.03}) is obtained by setting $c=0$, which is an allowed value for $c$. 
Following \cite{bershtein.foda}, the result can be re-written in the simpler form 
\begin{equation}
\label{i.i+n.04}
Y_{i+n,  \, \rrho} - Y_{i,  \, \rrho - 1 + \sum_{j=0}^{n-1} s_{i+j}} 
\geq                           1 - \sum_{j=0}^{n-1} r_{i+j}
\end{equation}

\subsubsection{Comparing the $\{Y_i, Y_{i+n} \}$ conditions and the $n$ adjacent 
$\{Y_i, Y_{i+1} \}$ conditions}
Since the row-lengths of a partition are weakly-decreasing, condition (\ref{i.i+n.04}) is 
satisfied if
\begin{equation}
\label{comparison.01}
Y_{i+n,  \, \rrho} - Y_{i,  \, \rrho - n + \sum_{j=0}^{n-1} s_{i+j}}
\geq                           1 - \sum_{j=0}^{n-1} r_{i+j},
\end{equation}
\noindent where $n > 1$, which is satisfied if 
\begin{equation}
\label{comparison.02}
Y_{i+n,  \, \rrho} - Y_{i,  \, \rrho - n + \sum_{j=0}^{n-1} s_{i+j}}
\geq                           n - \sum_{j=0}^{n-1} r_{i+j},
\end{equation}
\noindent which is condition (\ref{n.i.i+1.03}). Thus, the $n$ adjacent $\{Y_i, Y_{i+1} \}$ conditions 
(\ref{n.i.i+1.03}), which follow from the $\{Y_i, Y_{i+1}\}$ conditions (\ref{i.i+1.06}), are stronger 
than the $\{Y_i, Y_{i+n}\}$ conditions (\ref{i.i+n.04}), and it is sufficient to impose the underlying 
$\{Y_i, Y_{i+1} \}$ conditions (\ref{i.i+1.06}) to eliminate the zeros in the $\{Y_i, Y_{i+n}\}$. 

\subsection{Conditions from $\{Y_i, Y_{i-n} \}$}
\label{i.i-n}
Each of these products, for $i = 1, \cdots, N$, $n > 0$, vanishes if it contains a factor that satisfies 

\begin{multline}
\label{i.i-n.01}
E[ x_i, x_{i-n}, Y_i, Y_{i-n}, \square]
=
\\
\ll x^+_i - x^+_{i-n} + A^{+}_{\square, i  } \rr \, \apos
+
\ll x^-_i - x^-_{i-n} - L^{ }_{\square, i-n} \rr \, \aneg
=
\\
\ll   A^{+}_{\square, i  } + \sum_{j=1}^n r_{i-j} \rr \, \apos
+
\ll - L^{ }_{\square, i-n} + \sum_{j=1}^n s_{i-j} \rr \, \aneg
=
0,
\end{multline}

\noindent which leads to the conditions

\begin{equation}
\label{i.i-n.02}
  A^{ }_{\square, i  } = - 1 + c\   p - \sum_{j=1}^n r_{i-j}
\quad 
- L^{ }_{\square, i-n} =       c\ \pp - \sum_{j=1}^n s_{i-j}, 
\end{equation}

\noindent where $c$ remains to be determined. Following the same arguments used in 
{\bf \ref{i.i+1}}, $c$ must be a non-zero positive integer. In other words, conditions 
(\ref{i.i-n.02}) are possible for $c = 1, 2, \cdots$, $\square \in Y_i$ and 
$\square \not \in Y_{i-n}$.
\subsubsection{From two zero-conditions to one non-zero-condition} 
\label{translating.03}
Following paragraphs {\bf \ref{from.2.to.1}} and {\bf \ref{1.not.to.vanish}}, the two 
zero-conditions in (\ref{i.i-n.02}) can be translated to one non-zero-condition, 
\begin{equation}
\label{i.i-n.03}
Y^{\intercal}_{i-n, \, \ssigma} - Y^{\intercal}_{i, \, \ssigma - 1 + c\,   p - \sum_{j=1}^n r_{i-j}} 
\geq                                                       1 - c\, \pp + \sum_{j=1}^n s_{i-j} 
\end{equation}
\subsubsection{The strongest condition} 
Following the arguments in paragraph {\bf \ref{the.strongest.condition}}, the strongest version
of condition (\ref{i.i+n.03}) is obtained by setting $c=1$, which is an allowed value for $c$.
Following \cite{bershtein.foda}, the result can be re-written in the simpler form
\begin{equation}
\label{i.i-n.06}
Y_{i-n,  \, \rrho} - Y_{i,  \, \rrho - 1 + \pp - \sum_{j=1}^n s_{i-j}} 
\geq                           1 -   p + \sum_{j=1}^n r_{i-j} 
\end{equation}

\subsubsection{Comparing the conditions from $\{Y_i, Y_{i-n}\}$ and from $n$-adjacent 
$\{Y_i, Y_{i-1}\}$}
Using the $N$-periodicity of the partitions $Y_i$, $i \in \ZZ$, as well as the sum 
conditions (\ref{sum.conditions}), we can re-write (\ref{i.i-n.06}) as

\begin{equation}
\label{comparison.03}
Y_{i+N-n,  \, \rrho} - Y_{i, \ll \rrho - 1 + \sum_{j=n+1}^N s_{i-j} \rr}
\geq                     \ll         1 - \sum_{j=n+1}^N r_{i-j} \rr, 
\end{equation}

\noindent then using the $N$-periodicity of the integers $r_i$ and $s_i$, $i \in \ZZ$, we re-write 
(\ref{comparison.03}) as

\begin{equation}
Y_{i+N-n, \rrho} - Y_{i, \ll \rrho - 1 + \sum_{j=0}^{N-n-1} s_{i+j} \rr}
\geq                     \ll         1 - \sum_{j=0}^{N-n-1} r_{i+j} \rr,
\end{equation}

\noindent which is identical to the conditions (\ref{i.i+n.04}), upon a trivial change of labels. 
Thus, the conditions from $\{Y_i, Y_{i+1} \}$ (\ref{i.i+1.06}) 
are stronger than the conditions from  $\{Y_i, Y_{i+n} \}$ (\ref{i.i-n.06}), and it is sufficient 
to impose the former to eliminate the zeros in $\{Y_i, Y_{i-n}\}$, $n < 0$. 

\subsection{Conditions from $\{Y_i, Y_{i+1} \}^{\prime}$}
\label{conjugate.i.i+1}
Each of these products, for $i = 1, \cdots, N$, vanishes if it contains a factor that satisfies 

\begin{multline}
\label{conjugate.i.i+1.01}
- \apos - \aneg + E[ x_i, x_{i+1}, Y_i, Y_{i+1}, \square]
=
\\
\ll x^+_i - x^+_{i+1} + A^{+}_{\square, i  } - 1 \rr \, \apos
+
\ll x^-_i - x^-_{i+1} - L^{ }_{\square, i+1} - 1 \rr \, \aneg
=
\\
\ll - r_i + A^{ }_{\square, i  } \rr \, \apos
+
\ll - s_i - L^{+}_{\square, i+1} \rr \, \aneg
=
0,
\end{multline}

\noindent which, using (\ref{from.x.to.r.and.s}), leads to the conditions
\begin{equation}
\label{conjugate.i.i+1.02}
  A^{ }_{\square, i  } = r_i     + c\   p, 
\quad 
- L^{ }_{\square, i+1} = s_i + 1 + c\ \pp,  
\end{equation}

\noindent where $c$ remains to be determined. Following the same arguments used in
{\bf \ref{i.i+1}}, $c$ must be a non-negative integer. In other words, conditions 
(\ref{conjugate.i.i+1.02}) are possible for $c = 0, 1, 2, \cdots$, $\square \in Y_i$ 
and $\square \not \in Y_{i+1}$.
\subsubsection{From two zero-conditions to one non-zero-condition} 
\label{translating.04}
Following paragraphs {\bf \ref{from.2.to.1}} and {\bf \ref{1.not.to.vanish}}, the two 
zero-conditions in (\ref{conjugate.i.i+1.02}) can be translated to one non-zero-condition, 
\begin{equation}
\label{conjugate.i.i+1.03}
Y^{\intercal}_{i+1, \, \ssigma} - Y^{\intercal}_{i, \, \ssigma + r_i + c\, p} \geq - s_i - c\, \pp 
\end{equation}
\subsubsection{The stronger condition} 
\label{the.stronger.condition.01}
Since the row-lengths of a partition are weakly decreasing, condition (\ref{conjugate.i.i+1.03}) is 
satisfied if 

\begin{equation}
\label{conjugate.i.i+1.04}
Y^{\intercal}_{i+1, \, \ssigma} - Y^{\intercal}_{i, \, \ssigma + r_i} \geq - s_i - c\, \pp, 
\end{equation}

\noindent which is the case if 

\begin{equation}
\label{conjugate.i.i+1.05}
Y^{\intercal}_{i+1, \, \ssigma} - Y^{\intercal}_{1, \, \ssigma + r_i} \geq - s_i
\end{equation}

\noindent Thus, we should set $c=0$, which is an allowed value for $c$, in (\ref{conjugate.i.i+1.03}), 
and following \cite{bershtein.foda}, re-write it in the simpler form 

\begin{equation}
\label{conjugate.i.i+1.06}
Y_{i+1,  \, \rrho} - Y_{i,  \, \rrho + s_i} \geq - r_i 
\end{equation}

\subsubsection{Comparing conditions} Using the arguments of paragraph {\bf \ref{the.strongest.condition}}, 
one finds that conditions (\ref{i.i+1.06}) are stronger than conditions (\ref{conjugate.i.i+1.06}). Thus 
conditions (\ref{i.i+1.06}) that eliminate the zeros in 
$\{Y_i, Y_{i+1}\}$ are sufficient to eliminate the zeros in $\{Y_i, Y_{i+1}\}^{\prime}$.

\subsection{Conjugate products leads to weaker conditions} It is straightforward to see that all remaining 
conjugate products lead to conditions that are weaker than those of the corresponding products. The reason 
is that every $\{Y_i, Y_j\}^{\prime}$ is related to the corresponding product 
$\{Y_i, Y_j\}$ by replacing each elementary factor $E[x_i, x_j, Y_i, Y_j, \square]$ in 
$\{Y_i, Y_j\}$ by a factor $-\apos - \aneg + E[x_i, x_j, Y_i, Y_j, \square]$, up to an overall 
minus sign. As can be seen, by comparing the expressions in subsection {\bf \ref{i.i+1}} to the 
corresponding expressions in this subsection, this amounts to changing

\begin{equation}
A^{+}_{\square, i} \rightarrow A^{ }_{\square, i}, 
\quad
L^{ }_{\square, j} \rightarrow L^{+}_{\square, j},  
\end{equation}

\noindent where we have used $Y_i$ and $Y_j$ for generality. This leads to changing the final expressions 
for the non-zero conditions, 

\begin{equation}
r_i \rightarrow r_i + 1, 
\quad
s_i \rightarrow s_i + 1, 
\end{equation}

\noindent which, following the arguments in paragraph {\bf \ref{the.strongest.condition}}, leads to weaker 
conditions. In particular, the conditions obtained from 
$\{Y_i, Y_{i+n}\}^{\prime}$ and from
$\{Y_i, Y_{i-n}\}^{\prime}$ 
are weaker than those discussed in subsections {\bf \ref{i.i+n}} and {\bf \ref{i.i-n}}, respectively.

\subsubsection{The conditions from $\{Y_i, Y_{i+1}\}$ are sufficient} From the above, 
we conclude that the $N$-Burge conditions (\ref{i.i+1.06}), which we recall for convenience,

\begin{equation}
\label{i.i+1.06.recalled}
\boxed{
Y_{i+1,  \, \rrho} - Y_{i,  \, \rrho + \ll s_i - 1 \rr} \geq - \ll r_i - 1 \rr
}
\end{equation}

\noindent are sufficient to eliminate all zeros in all products in $z_{\it den}$. 

\subsubsection{Cylindric partitions. $N$-Burge conditions} 
\label{cylindric.N.partitions}
The conditions (\ref{i.i+1.06}) form a special 
case of those that were introduced and studied in \cite{gessel.krattenthaler}. A set of $N$ partitions
that satisfy such conditions are called {\it \lq cylindric partitions\rq} in \cite{gessel.krattenthaler}.
They have appeared in this specific form in \cite{feigin.feigin.jimbo.miwa.mukhin.02}.  

%SECTION.05
\section{$\cW_3$ minimal model conformal blocks from AGT with restricted Young diagrams}
\label{w.3.minimal.models}

{\it We check the validity of the expression in equation (\ref{restricted.sum}) by computing a non-trivial 
conformal block in a $\cW_3$ minimal model. To do that, we consider a $\cW_3$ conformal block that is known 
to satisfy a third-order ordinary differential equation of Pochammer type, that is solved in terms of ${}_3F_2$ 
Hypergeometric functions \cite{fateev.litvinov.2007}. To be reasonably self-contained, we outline the derivation 
of this differential equation in some detail.
} 

\subsection{A family of holomorphic 4-point functions}
\label{correlation.function}

Consider the $\cW_3$ holomorphic 4-point functions 

\begin{equation}
\label{holomorphic.4.point.function}
F\{z\} = 
\langle \, \prod_{i=0}^3 \cO_{\vP_i} (z_i) \, \rangle,
\end{equation}

\noindent where $\{z\} = \{z_0, \cdots, z_3\}$ is a set of four points on the Riemann sphere, 
$\cO_{\vP_i} (z_i)$, $i=0, \cdots, 3$, is a vertex operator insertion at $z_i$. $\cO_{\vP_i}$ 
is a vertex operator that inserts a $\cW_3$ highest weight state labeled by the charge vector 
$\vP_i$. At this point, the charge vectors $\vP_i$, $i=0, \cdots, 3$, could be any vectors in 
the weight lattice of the Lie algebra $A_2$, spanned by the fundamental weight vectors 
$(\vomega_1, \vomega_2)$, 

\begin{equation}
\label{choice.1}
\vP_0 = c_{1, i} \ \vomega_1 + c_{2, i} \ \vomega_2, 
\end{equation}

\noindent where $c_{1, i}, c_{2, i} \in \RR$. 

\subsection{Specialising the 4-point functions}
For the purposes of this section, we chose to keep 
$\vP_0$ and $\vP_3$ arbitrary, and set $\vP_1$ and $\vP_2$ to point in the direction 
of $\vomega_1$ only, such that

\begin{equation}
\label{choice.2}
\vP_1 = -b \, \vomega_1, \quad
\vP_2 =  a \, \vomega_1, 
\end{equation}

\noindent where $b$ is the parameter that determines the Virasoro central charge, see equation 
(\ref{central.charge}), while $a \in \RR$ is arbitrary. Using global conformal invariance 
\cite{zam.zam}, the holomorphic 4-point function in equation (\ref{holomorphic.4.point.function}), 
with the charge vectors chosen as in equations (\ref{choice.1}) and (\ref{choice.2}), can be 
written in the form 
\footnote{\,
The subscripts $\{0, 1  2, 3\}$ that we use  to label the points of the 4-point function on 
the Riemann sphere, correspond to 
the subscripts $\{2, 1, 3, 4\}$         used to label the same points in \cite{zam.zam}.
}

\begin{multline}
\label{f.b}
F \{z\} =
\, z_{31}^{ -2 \Delta_1}
\, z_{30}^{\ll \Delta_1 + \Delta_2 - \Delta_3 - \Delta_0 \rr}
\, z_{32}^{\ll \Delta_1 + \Delta_0 - \Delta_2 - \Delta_3 \rr}
\, z_{20}^{\ll \Delta_3 - \Delta_1 - \Delta_2 - \Delta_0 \rr} 
\\
\cB \ll z| \vP_0, \vP_{\, \textit{int}}, \vP_3 |-b, a \rr,
\end{multline}

\noindent where $\cB$ is the factor in $F\{z\}$ that depends only on $z$
\footnote{\, 
At this stage, we are working in the context of generic $\cW_N$ models.In particular, we did not 
choose the parameter $b$ in equation (\ref{choice.2}) such that we obtain a minimal $\cW_N$ model. 
Since we have specialised to $\cW_3$, that is $N=3$, and we focus on 4-point functions, that is 
$n=1$, once we choose the parameters in $F\{z\}$ to be those of a minimal model labeled by the 
coprimes $p$ and $\pp$, $\cB$ in equation (\ref{f.b}) becomes $\cB^{\, p, \pp}_{3, 1}$ in the 
notation of equation (\ref{conformal.block}).  
}, 
the projective invariant cross-ratio of the coordinates, 

\begin{equation}
\label{cross.ratio}
z=\frac{z_{10} z_{23}}{z_{13} z_{20}}, \quad z_{ij}=z_i-z_j, 
\end{equation}

\noindent and $\Delta_i$ is the conformal dimension of $\cO_{\vP_i} (z_i)$, see equation
$(\ref{conformal.dimension})$. 

The relevant factor on the right hand side of equation (\ref{f.b}) is 
$\cB [z| \vP_0, \vP_{\, \textit{int}}, \vP_3 |-b, a]$, all parameters of which are 
initial data that specify the 4-point function that we wish to compute, except 
$\vP_{\, \textit{int}}$ which remains to be determined. 
Following \cite{fateev.litvinov.2007, mironov.morozov}, 
$\cB [z| \vP_0, \vP_{\, \textit int}, \vP_3|-b, a]$ satisfies a third-order ordinary 
differential equation with respect to $z$. Requiring that $\cB$ satisfies this 
differential equation, determines the three possible values of the charge vector $\vP_{\, \textit int}$ 
of the $\cW_3$ highest weight representation that flows in the internal channel of the conformal 
block. In the rest of this section, we derive, this equation in six steps. 

\subsubsection{Step 1.}
The $\cW_3$ Verma module of highest weight vector $(-2 b\; \vomega_1 - b^{-1}\; \vomega_2)$, associated 
to the vertex operator $\cO_{\vP_1}$, contains a $\cW_3$ null-state at level $3$. This implies that
$F\{z\}$ in equation (\ref{f.b}) satisfies the null state condition

\begin{multline}
\label{null.state.condition}
\ll 
W_{-3}^{(1)}-\frac{16 w_1}{ \Delta_1 (\Delta_1 + 1) (5 \Delta_1 +1)} \ll L_{-1}^{(1)} \rr^3 + 
\right.
\\
\left.
\frac{12 w_1}{\Delta_1 (5 \Delta_1 + 1)} L_{-1}^{(1)} L_{-2}^{(1)}+\frac{3w_1(\Delta_1-3)}{2 \Delta_1 (5 \Delta_1 + 1)}
L_{-3}^{(1)} 
\rr 
F \{z\} = 0,
\end{multline}

\noindent where the generators $L_{-m}^{(1)}$, $m=1, 2, 3$, and $W_{-3}^{(1)}$ act on $\cO_{\vP_1} (z_1)$. 
We need to express this action in terms of the differential operator action.   

\subsubsection{Step 2.} 
We use the $\cW_3$ Ward identity

\begin{equation}
\label{w3.ward.identity}
W_{-3}^{(1)} F \{z\} =   \sum_{\substack{j=0 \\ j \neq 1}}^3 
\ll
\frac{w_j         }{(z_i-z_j)^3}+
\frac{W_{-1}^{(j)}}{(z_i-z_j)^2}+ 
\frac{W_{-2}^{(j)}}{(z_i-z_j)  }
\rr          F \{z\},
\end{equation}

\noindent to express the action of $W_{-3}^{(1)}$ on $F\{z\}$ in equation (\ref{null.state.condition}),
in terms of the action of the six lower-degree generators $W_{-1}^{(j)}$, and $W_{-2}^{(j)}$, $j = 0, 2, 3$, 
which act on the other three vertex operators in $F\{z\}$.

\subsubsection{Step 3.} We use the five $\cW_3$ Ward identities \cite{fateev.litvinov.2007}, 

\begin{align}
&\sum_{j=0}^3           \, W_{-2}^{(j)}                                        =0,\\
&\sum_{j=0}^3 \ll z_j   \, W_{-2}^{(j)} +         \, W_{-1}^{(j)}             \rr =0,\\
&\sum_{j=0}^3 \ll z_j^2 \, W_{-2}^{(j)} + 2 z_j   \, W_{-1}^{(j)}+        w_j \rr =0,\\
&\sum_{j=0}^3 \ll z_j^3 \, W_{-2}^{(j)} + 3 z_j^2 \, W_{-1}^{(j)}+3 z_j   w_j \rr =0,
\end{align}
\begin{align}
&\sum_{j=0}^3 \ll z_j^4 W_{-2}^{(j)} + 4 z_j^3 W_{-1}^{(j)}+6 z_j^2 w_j \rr =0.
\end{align}
\noindent to express the five generators $W_{-1}^{(0)}$, $W_{-2}^{(0)}$, $W_{-1}^{(3)}$, $W_{-2}^{(3)}$ 
and $W_{-2}^{(2)}$ in terms of the three generators $W_{-1}^{(1)}$, $W_{-2}^{(1)}$, and $W_{-1}^{(2)}$. 

\subsubsection{Step 4.} We use the fact that there are null-states at level-1 and level-2 in the Verma 
module with highest weight vector $(-2 b\; \vomega_1 - b^{-1}\; \vomega_2)$, and that there is a null-state 
at level-1 in the Verma module with highest weight state vector $(-(b+a)\; \vomega_1 - b^{-1}\; \vomega_2)$, 
to obtain the relations

\begin{align}
& W_{-1}^{(1)} F \{z\} = \frac{3 w_1}{2 \Delta_1} \partial_{z_1} F \{z\}
\label{WWard.1}
\\
& W_{-2}^{(1)} F \{z\} = 
\ll \frac{12 w_1             }{\Delta_1 (5 \Delta_1 + 1)} \partial_{z_1}^2 - 
    \frac{ 6 w_1 (\Delta_1+1)}{\Delta_1 (5 \Delta_1 + 1)} L_{-2}^{(1)} \rr F \{z\} 
\label{WWard.2}
\\
& W_{-1}^{(2)} F \{z\} =\frac{3w_2}{2 \Delta_2}\partial_{z_2} F \{z\},
\label{WWard.3}
\end{align}

\subsubsection{Step 5.} 
To express the action of the Virasoro generators $L_{-m}^{(1)}$, $m=1, 2, 3$, in above equations, 
in terms of differential operators, we use the conformal Ward identity

\begin{equation}
\label{conformal.ward.identity}
L_{-m}^{(i)} F \{z\} = - \sum_{\substack{j=0 \\ j \neq i}}^3 
\ll 
\frac{(m-1)  \Delta_j}{(z_i-z_j)^m}+
\frac{\partial_{z_j}}{(z_i-z_j)  }
\rr          F \{z\}
\end{equation}

\subsubsection{Step 6.}

To obtain an ordinary differential equation for the 4-point function $F\{z\}$, it is convenient to work 
in terms of $\cB^{\, \textit simple}$, that is defined in terms of $\cB$, the projective-invariant 
component of $F\{z\}$ in equation (\ref{f.b}), 

\begin{multline}
\label{b.b.simple}
\cB^{\, \textit simple} \ll z| \vP_0, \vP_{\text{int}}, \vP_3 | -b, a \rr = 
\\
                        \ll  z^{-\frac{b (2 \ell_1 + \ell_2) }{3}} \, (1-z)^{-\frac{3+3 b^2-b a}{3}} \rr 
\cB^{               }   \ll z| \vP_0, \vP_{\text{int}}, \vP_3 | -b, a \rr,
\end{multline}

\begin{equation}
\label{ell}
\ell_i = \langle \vP_0 | \valpha_i \rangle + \ll b + \frac{1}{b} \rr, \quad i = 1, 2
\end{equation}

\subsection{The differential equation}

Following \cite{fateev.litvinov.2007}, we combine the above equations and find that $\cB^{\, \textit simple}$ 
satisfies the Pochhammer generalised hypergeometric differential equation 

\begin{equation}
\label{pochhamer.1}
\cD_z \ \cB^{\, \textit simple} \ll z| \vP_0, \vP_{\text{int}}, \vP_3 | -b, a \rr = 0,
\end{equation}

\begin{equation}
\label{difeq}
\\
\cD_z \equiv 
z \ll \partial_{z} + A_1     \rr \ll \partial_z + A_2     \rr \ll \partial_z + A_3 \rr 
-
  \ll \partial_{z} + B_1 - 1 \rr \ll \partial_z + B_2 - 1 \rr     \partial_z, 
\\
\end{equation}

\begin{multline}
A_i = \frac{b^2 + 3 - b \; a}{3} + b \, \langle \vP_0 | \vh_1  \rangle 
                                 + b \, \langle \vP_3 | \vh_i  \rangle, \quad i = 1, 2, 3,
\\
B_1 = 1                        +b \, \langle \vP_0 | \valpha_1     \rangle, \quad 
B_2 = 1                        +b \, \langle \vP_0 | \valpha_1+ \valpha_2 \rangle 
\end{multline}

Recalling that $\vP_{\, \textit int}$ is the only undetermined parameter in $\cB^{\, \textit{simple}}$, 
and requiring that $\cB^{\, \textit simple}$ satisfies equation (\ref{pochhamer.1}), implies that there 
are at most three possible values for $\vP_{\, \textit int}$, and correspondingly, at most three possible 
$\cW_3$ highest weight modules are allowed to propagate in the intermediate channel of the 4-point function. 
The solution of equation (\ref{pochhamer.1}) is known to be a hypergeometric function of type ${}_3 F_2$
\cite{fateev.litvinov.2007}.

The differential equation that $\cB$ satisfies is obtained by composing the factor on the right hand 
side of equation (\ref{b.b.simple}) and $\cD_z$ in equation (\ref{pochhamer.1}), 

\begin{equation}
\label{pochhamer.2}
\ll 
\cD_z 
\circ 
\ll  z^{-\frac{b (2 \ell_1 + \ell_2) }{3}} \, (1-z)^{-\frac{3+3 b^2-b a}{3}} \rr
\rr 
\,
\cB \ll z| \vP_0, \vP_{\text{int}}, \vP_3 | -b, a \rr = 0
\end{equation}

\noindent Proposing the $z$ expansion

\begin{equation}
\cB \ll z| \vP_0, \vP_{\text{int}},  \vP_3 | -b, a \rr = z^\gamma \ll 1 + \cO(z) \rr,
\end{equation}

\noindent one looks for the possible values of $\gamma$ that satisfy equation (\ref{pochhamer.2})
to leading order. The values $\gamma_i$, $i = 1, 2, 3$, that solve the third-order algebraic 
characteristic equation can be written as 

\begin{equation}
\gamma_i = \Delta_{\vP_{\, \textit int, i}} - \Delta_0 - \Delta_1, 
\end{equation}

\noindent where $\vP_{\, \textit int, i}$, $i = 1, 2, 3$, are the charge vectors of the $\cW_3$ 
highest weight modules that are allowed to floow in the internal channel. The values of 
$\vP_{\textit int,  i}$ that we obtain are

\begin{equation}
\label{thecase} 
\vP_\textit{int,1} = \vP_0 - b \; \vomega_1,               \quad 
\vP_\textit{int,2} = \vP_0 + b \; \vomega_2,               \quad 
\vP_\textit{int,3} = \vP_0 + b \; \vomega_1 - b \; \vomega_2, 
\end{equation}
 
\noindent where the charge vector $\vP_0$, which is an arbitrary vector in the $A_2$ weight lattice, 
the parameter $b$ that determines the Virasoro central charge, and the arbitrary real parameter $a$,
are the external data that specify $\cB$. 

In the following, we focus on the solution of (\ref{pochhamer.2}) that corresponds to the internal 
channel that carries the $\cW_3$ module with highest weight vector $\vP_\textit{\, int, 1}$. In this 
specific case, we obtain

\begin{equation}
\label{G.s.0}
\cB \ll z| \vP_0, \vP_0 - b, \vP_3 | -b, a \rr  =
    \ll  z^{\frac{b (2 \ell_1 + \ell_2) }{3}} \, (1-z)^{\frac{3+3 b^2-b a}{3}} \rr
{}_3 F_2 \ll A_1, A_2, A_3; B_1, B_2; z \rr
\end{equation}

\subsection{Minimal $\cM^{\, p, \, \pp}_3$ models}

The above results, obtained from general properties of $\cW_3$ algebra, are valid for all values of the 
Virasoro central charge $c$. To check equation (\ref{restricted.sum}), we specialize to the 
$\cW_3$ minimal models $\cM^{\, p, \, \pp}_3$, where $p$ and $\pp$ are coprime integers that 
satisfy $3 \leq p < \pp$. To do this, we set 

\begin{equation}
b      \to i \apos, 
\quad 
b^{-1} \to i \aneg, 
\quad 
\apos =    \ll \frac{\pp}{p} \rr^{\frac12},
\quad
\aneg =  - \ll \frac{p}{\pp} \rr^{\frac12}
\end{equation}

\noindent so that, from equation (\ref{c.p.p.prime}), we obtain the Virasoro central charge 

\begin{equation}
\label{c.p.p.prime.2}
c_3^{\, p, \, \pp} = 2 \ll 1 - 12 \, \alpha_0^2 \rr, \quad \alpha_0 = \apos + \aneg
\end{equation}

\noindent Further, we associate each vertex operator $\cO_{\vP}$ to a highest weight vector 
$\vP_{\, \vec r \, \vec s}$, where $\vec r = \{r_1, r_2\}$, $\vec s = \{s_1, s_2\}$, such that 
$r_1$, $r_2$, $s_1$ and $s_2$ are integers that satisfy 

\begin{equation}
1 \leq  \ll \sum_{i=1}^2 r_i \rr \leq  p, \quad
1 \leq  \ll \sum_{i=1}^2 s_i \rr \leq \pp
\end{equation}

In the sequel, we simplify the notation by writing the charge vector 
$\vP_{\, \vec r,  \, \vec s}$ as 
$\vP_{\, r_1, r_2; s_1, s_2}$, 
and the corresponding Virasoro and $\cW_3$ eigenvalues  
$\Delta_{\vP_{\, \vec r, \, \vec s}}$ and 
$     w_{\vP_{\, \vec r, \, \vec s}}$ as 
$\Delta_{\, \vec r, \, \vec s}$ and 
$     w_{\, \vec r, \, \vec s}$.

\subsection{Checking the modified AGT expression}
\label{checking}
We want to check equation (\ref{restricted.sum}) for the non-trivial conformal block, 
computed in (\ref{G.s.0}), after specializing the parameters to minimal model ones.
We choose $N=3$, $p=8$ and $\pp=9$, and consider the function (\ref{G.s.0}) for the unitary minimal model 
$\cM^{\, 8, \, 9}_3$, with 

\be 
\vP_0 = \vP_{11; 12}, \quad 
\vP_3 = \vP_{11; 21}, \quad 
  a   = -b,           \quad
\vP_{\textit int, 1} = \vP_0 - b \, \vomega_1 = \vP_{21; 12}
\ee

\noindent and compare with the result obtained by applying equation (\ref{restricted.sum}) to 
$\cB^{\, 8, \, 9}_{3, 1} [z| \vP_0 \, \vP_{\textit int, 1} \, \vP_3 |-b, -b]$, see equation 
(\ref{conformal.block}). The $\cW_3$ irreducible highest weight module that flows in the 
intermediate channel in this case is characterised by 
$\{p, \pp, r_1, r_2, s_1, s_2\} = $
$\{8,   9,   2,   1,   1,   2\}$, and the triples of Young diagrams that are allowed by the 
3-Burge conditions, in this case, for $| Y | = 0, 1, 2, 3$ and $4$, where $| Y |$ is 

\begin{equation}
| Y | = \sum_{i=1}^3 | Y_i | 
\end{equation}

\noindent and $| Y_i |$, $i = 1, 2, 3$, is the number of cells in the $i$-th Young diagram, 
are

\be
\nonumber
\ba{c}
\dps
| Y | = 0: 
\ll \varnothing, \varnothing, \varnothing \rr, 
\quad
| Y | = 1: 
\ll \varnothing, \varnothing, \tableau{1} \rr,
\ll \varnothing, \tableau{1}, \varnothing \rr,
\ll \tableau{1}, \varnothing, \varnothing \rr
\ea
\ee

\be
\nonumber
\ba{l}
\dps
| Y | = 2: 
\ll \varnothing,   \varnothing, \tableau{2} \rr,
\ll \varnothing,   \tableau{1}, \tableau{1} \rr,
\ll \varnothing,   \tableau{2}, \varnothing \rr,
\ll \tableau{1},   \varnothing, \tableau{1} \rr,
\ll \tableau{1},   \tableau{1}, \varnothing \rr,
\\
\\
\hspace{16mm}
\ll \tableau{2},   \varnothing, \varnothing \rr,
\ll \tableau{1 1}, \varnothing, \varnothing \rr
\ea
\ee

\be
\nonumber
\ba{l}
\dps
| Y | = 3: 
\ll \varnothing,\varnothing,\tableau{3}   \rr,
\ll \varnothing,\tableau{1},\tableau{2}   \rr,
\ll \varnothing,\tableau{1},\tableau{1 1} \rr,
\ll \varnothing,\tableau{2},\tableau{1}   \rr,
\ll \varnothing,\tableau{3},\varnothing   \rr,
\\
\\
\hspace{16mm} 
\ll \tableau{1},\varnothing,\tableau{2}   \rr,
\ll \tableau{1},\tableau{1},\tableau{1}   \rr,
\ll \tableau{1},\tableau{2},\varnothing   \rr,
\ll \tableau{1},\tableau{1 1},\varnothing \rr,
\ll \tableau{2},\varnothing,\tableau{1}   \rr,
\\
\\
\hspace{16mm} 
\ll \tableau{1 1},\varnothing,\tableau{1} \rr,
\ll \tableau{2},\tableau{1},\varnothing   \rr,
\ll \tableau{1 1},\tableau{1},\varnothing \rr,
\ll \tableau{3},\varnothing,\varnothing     \rr,
\ll \tableau{2 1},\varnothing,\varnothing   \rr,
\\
\\
\hspace{16mm}
\ll \tableau{1 1 1},\varnothing,\varnothing \rr
\ea
\ee

\be
\nonumber
\ba{l}
\dps
| Y | = 4:
\ll \varnothing,\varnothing,\tableau{4}   \rr,
\ll \varnothing,\tableau{1},\tableau{3}   \rr,
\ll \varnothing,\tableau{1},\tableau{2 1} \rr,
\ll \varnothing,\tableau{2},\tableau{2}   \rr,
\ll \varnothing,\tableau{2},\tableau{1 1} \rr,
\\
\\
\hspace{16mm} 
\ll \varnothing,\tableau{3},\tableau{1}   \rr,
\ll \varnothing,\tableau{4},\varnothing   \rr,
\ll \tableau{1},\varnothing,\tableau{3}   \rr,
\ll \tableau{1},\tableau{1},\tableau{2}   \rr,
\ll \tableau{1},\tableau{1},\tableau{1 1} \rr,
\\
\\
\hspace{16mm} 
\ll \tableau{1},\tableau{2},\tableau{1}   \rr,
\ll \tableau{1},\tableau{1 1},\tableau{1} \rr,
\ll \tableau{1},\tableau{3},\varnothing   \rr,
\ll \tableau{1},\tableau{2 1},\varnothing \rr,
\ll \tableau{2},\varnothing,\tableau{2}   \rr,
\\
\\
\hspace{16mm} 
\ll \tableau{1 1},\varnothing,\tableau{2} \rr,
\ll \tableau{2},\tableau{1},\tableau{1}   \rr,
\ll \tableau{1 1},\tableau{1},\tableau{1} \rr,
\ll \tableau{2},\tableau{2},\varnothing     \rr,
\ll \tableau{2},\tableau{1 1},\varnothing   \rr,
\\
\\
\hspace{16mm} 
\ll \tableau{1 1},\tableau{2},\varnothing   \rr,
\ll \tableau{1 1},\tableau{1 1},\varnothing \rr,
\ll \tableau{3},\varnothing,\tableau{1}     \rr,
\ll \tableau{2 1},\varnothing,\tableau{1}   \rr,
\ll \tableau{1 1 1},\varnothing,\tableau{1} \rr,
\ea
\ee

\be
\nonumber
\ba{l}
\dps
\hspace{16mm} 
\ll \tableau{3},\tableau{1},\varnothing     \rr,
\ll \tableau{2 1},\tableau{1},\varnothing   \rr,
\ll \tableau{1 1 1},\tableau{1},\varnothing \rr,
\ll \tableau{4},\varnothing,\varnothing     \rr,
\ll \tableau{3 1},\varnothing,\varnothing   \rr,
\\
\\
\hspace{16mm} 
\ll \tableau{2 2},\varnothing,\varnothing     \rr, 
\ll \tableau{2 1 1},\varnothing,\varnothing   \rr,
\ll \tableau{1 1 1 1},\varnothing,\varnothing \rr
\ea
\ee

Considering the contribution of the allowed Young diagrams only, we obtain 

\begin{multline}
\cB^{\, 8, \, 9}_{3, 1}  \ll z| \vP_0 \vP_1^{(1)}\vP_3 |-b, -b \rr =  
\\
1 + \frac{    32}{    135} z  
  + \frac{   101}{    729} z^2  
  + \frac{ 64576}{ 649539} z^3  
  + \frac{124748}{1594323} z^4 + 
\\
\frac{  30730880}{  473513931} z^5 + 
\frac{ 970725028}{17433922005} z^6 +
\frac{4604400320}{94143178827} z^7 + \cO (z^8), 
\end{multline}

\noindent which coincides with ${}_3 F_2$ on the right hand side of equation \eqref{G.s.0}.
In other words, $\cB^{\, 8, \, 9}_{3, 1}$ coincides with $\cB$ in equation (\ref{f.b}) 
up to the normalisation factor  
$z^{-\frac{b (2 \ell_1 + \ell_2) }{3}}$
and the Heisenberg factor 
$(1-z)^{-\frac{3+3 b^2-b a}{3}}$. 

%SECTION.06
\section{$\cW_3$ minimal model characters from $3$-Burge partitions}
\label{wn.characters}
 
{\it We compare the characters of degenerate $\cW_3$ irreducible highest weight 
representations with the generating functions of triples of Young diagrams that 
obey 3-Burge conditions.}

\subsection{$\cW_3$ minimal model characters}

Expressions for the characters of degenerate irreducible $\cW_N$ highest weight representations 
were computed in \cite{frenkel.kac.wakimoto}. In the following, we specialise these expressions 
to the $N=3$ case, explain what the various terms are, how to evaluate them, then compute examples
of the characters in $q$-series form 
\footnote{\,
The notation used in this section is close to that in \cite{feigin.feigin.jimbo.miwa.mukhin.01}. 
}.

The $\cW_3$ minimal model character, labeled by two coprime integers $p$, $\pp$, such that 
$2 < p < \pp$, and dominant integral weight vectors $\bet$ and $\bxi$, of level-[$p$-3] 
and level-[$\pp$-3], respectively is 
\begin{multline}
\label{bar.02}
\chi^{\, p, \, \pp}_{\bets, \bxis} (q) = 
\frac{1}{\eta(q)^2}
\sum_{\sigma \in S_3} (-1)^{L_{\sigma}}
\sum_{r, s = - \infty}^{\infty}
q^{\frac{p \pp}{2}
\big\langle r \alpha_1 + s \alpha_2, \ r \alpha_1 + s \alpha_2 \big\rangle 
} 
\times
\\
q^{ 
\big\langle
    \pp \sigma \ll \sum_{i=0}^2 n_i \vomega_i \rr -  
      p        \ll \sum_{i=0}^2 m_i \vomega_i \rr, \  
      r \alpha_1 + s \alpha_2
\big\rangle
+
\big\langle
\sum_{i=0}^2 n_i \vomega_i - \sigma \ll \sum_{i=0}^2 n_i \vomega_i \rr, 
\sum_{i=0}^2 m_i \vomega_i
\big\rangle
}
\end{multline}

We need to explain what the various terms in equation (\ref{bar.02}) stand for, and how
to compute them. 
As mentioned above, $p$ and $\pp$ are coprime integers that satisfy $2 < p < \pp$. 
$\bet$ is a level-[$p$-3] dominant integral weight vector, and 
$\bxi$ is a level-[$\pp$-3] dominant integral weight vector. They are defined as
\begin{align}
&\bet=(n_0-1)\vomega_0+(n_1-1)\vomega_1+(n_2-1)\vomega_2, \quad n_0 + n_1 + n_2 = p, 
\\
&\bxi=(m_0-1)\vomega_0+(m_1-1)\vomega_1+(m_2-1)\vomega_2, \quad m_0 + m_1 + m_2 = \pp
\end{align}

\noindent $q$ is an indeterminate, and $\eta(q)$ is the Dedekind function 

\begin{equation}
\eta(q) = q^{1/24} \prod_{i=1}^{\infty} (1-q^i)
\end{equation}

\noindent $S_3$ is the symmetric group of degree 3, generated by the permutation 
operators $s_1$ and $s_2$,

\begin{equation}
\label{s3}
S_3 = \{1, s_1, s_2, s_1 s_2, s_2 s_1, s_1 s_2 s_1\}
\end{equation}

\noindent $L_{\sigma}$ is the length function of a permutation $\sigma$, that is, the minimal 
number of $S_3$ generators required to generate $\sigma$.  
Denoting the integral vector $\sum_{i=0}^{2} n_i \vomega_i$, $n_i \in \NN$, $i=0, 1, 2$, by 
$[n_0, n_1, n_2]$, the action of $\sigma$ on integral vectors in the $\widehat{A}_2$ weight 
lattice is 

\begin{eqnarray}
          1[n_0, n_1, n_2] &=& [n_0,                     n_1,          n_2],
\\
        s_1[n_0, n_1, n_2] &=& [n_0 +   n_1,           - n_1,          n_1 + n_2],
\nonumber
\\
        s_2[n_0, n_1, n_2] &=& [n_0 +   n_2,             n_1 + n_2,  - n_2],
\nonumber
\\
    s_1 s_2[n_0, n_1, n_2] &=& [n_0 + 2 n_2 +   n_1,   - n_1 - n_2,    n_1],
\nonumber
\\
    s_2 s_1[n_0, n_1, n_2] &=& [n_0 + 2 n_1 +   n_2,     n_2,        - n_1  -n_2],
\nonumber
\\
s_1 s_2 s_1[n_0, n_1, n_2] &=& [n_0 + 2 n_1 + 2 n_2,   - n_2,        - n_1]
\nonumber
\end{eqnarray}

\noindent The $\widehat{A}_2$ simple root vectors satisfy 

\begin{equation}
\label{root.inner.products}
\langle \valpha_1, \valpha_1 \rangle =
\langle \valpha_2, \valpha_2 \rangle = 2,  
\langle \valpha_1, \valpha_2 \rangle =-1
\end{equation}

\noindent The $\widehat{A}_2$ fundamental weight vectors satisfy 

\begin{equation}
\label{weight.inner.products}
\langle \vomega_0, \vomega_0 \rangle = \langle \vomega_0, \vomega_1 \rangle = \langle \vomega_0, \vomega_2 \rangle = 0, 
\langle \vomega_1, \vomega_1 \rangle = \langle \vomega_2, \vomega_2 \rangle = \frac23, 
\langle \vomega_1, \vomega_2 \rangle = \frac13,
\end{equation}

\noindent From 

\begin{equation}
\alpha_1 = - \vomega_0 + 2 \vomega_1 -   \vomega_2, \quad 
\alpha_2 = - \vomega_0 -   \vomega_1 + 2 \vomega_2,
\end{equation}

\noindent we have 

\begin{equation}
\langle \vomega_1,\alpha_1 \rangle = \langle \vomega_2, \alpha_2 \rangle = 1, 
\quad
\langle \vomega_1,\alpha_2 \rangle = \langle \vomega_2, \alpha_1 \rangle =0
\end{equation}

\noindent From the above equations, it is straightforward to show that  

\begin{multline}
\\
\chi^{\, p, \, \pp}_{\bets, \bxis} (q) =                               
                                  F \ll n_1,       m_1|  n_2,        m_2 \rr 
  - q^{n_1    m_1}                F \ll n_1,       m_1|  n_1 + n_2,  m_2 \rr 
\\
  - q^{ n_2   m_2}                F \ll n_1 + n_2, m_1 | -n_2,       m_2 \rr
  + q^{(n_1 + n_2) m_1 + n_2 m_2} F \ll n_1 - n_2, m_1 |  n_1,       m_2 \rr
\\
  + q^{(n_1 + n_2) m_2 + n_1 m_1} F \ll n_2,       m_1 | -n_1 - n_2, m_2 \rr
  - q^{(n_1 + n_2)(m_1 +    m_2)} F \ll n_2,       m_1 | -n_1,       m_2 \rr
\end{multline}

\be
F \ll x_1, x_2 | y_1, y_2 \rr = 
\frac{1}{\eta(q)^2} \sum_{r, s \in \ZZ} 
q^{p \pp (r^2 +s^2 -rs) + (\pp x_1 -p y_1)r +(\pp x_2 -p y_2)s}, 
\ee

\subsection{Examples}

We find the following $q$-series expansions 

\begin{multline}
\\
\chi^{3, 7}_{11 | 11} 
= 1 +         q^2 + 2 q^3 + 3 q^4 +  3 q^5 +  6 q^6 +  7 q^7 + 11 q^8 + 14 q^9 + 20 q^{10} + \cdots, 
\\
\chi^{3, 7}_{21 | 11} = \chi^{3, 7}_{11 | 21}
= 1 +   q + 2 q^2 + 3 q^3 + 5 q^4 +  7 q^5 + 11 q^6 + 14 q^7 + 21 q^8 + 28 q^9 + 39 q^{10} + \cdots,
\\
\chi^{3, 7}_{21 | 21} 
= 1 + 2 q + 3 q^2 + 5 q^3 + 8 q^4 + 11 q^5 + 17 q^6 + 24 q^7 + 34 q^8 + 47 q^9 + 64 q^{10} + \cdots,
\\
\chi^{3, 7}_{31 | 11} = \chi^{3, 7}_{11 | 31}
= 1 +   q + 3 q^2 + 3 q^3 + 6 q^4 +  8 q^5 + 13 q^6 + 17 q^7 + 25 q^8 + 33 q^9 + 47 q^{10} + \cdots
\end{multline}

Comparing the above expressions with those obtained from counting triples of Young diagrams that 
satisfy the 3-Burge conditions, we find that they coincide.

%SECTION.07
\section{Summary and comments}
\label{remarks}

{\it 
We propose a modified $\cW_N$ AGT prescription to allow one to compute conformal blocks 
$\cB^{\, p, \, \pp, \cH}_{N, n}$, from which one can extract $\cW_N$ minimal model conformal 
blocks $\cB^{\, p, \, \pp     }_{N, n}$. 
}

\subsection{$\cW_N$ AGT leads to ill-defined expressions in minimal model conformal blocks}
Applying the original, unmodified $\cW_N$ AGT correspondence to the minimal $\cW_N$ models, 
times contributions from a free boson, by setting the gauge theory mass and Coulomb parameters 
to minimal $\cW_N$ model values, and {\it leaving all else the same}, leads to ill-defined 
expressions in the form of zero divided by zero. 
The origin of these ill-defined expressions was explained in the context of $\cW_2$ in 
\cite{alkalaev.belavin, bershtein.foda}. We review it below.

\subsubsection{Norms and couplings of states that flow in channels}
The original prescription allows for all states in a specific $\cW_N$ Verma module to flow in 
each specific channel. The norms of these states appear in the denominators of Nekrasov's 
instanton partition function. Their coupling to other states are given by matrix elements
of the $\cW_N \! \times \cH$ algebra. These matrix elements appear as the factors in the 
numerators. 
When the central charge is non-minimal, there are no zero-norm states in the Verma module, all 
terms in the denominators are non-vanishing, the expressions are well defined regardless of 
whether the corresponding matrix elements that describe the couplings to other states are zero 
or not, and one obtains the correct result.
When the central charge is minimal, the situation is drastically different.

\subsubsection{The zeros in the denominators} 
When the central charge is minimal, there are zero-norm states in the Verma module. Including 
these states in the sums, one obtains zeros in the denominators. This is the origin of the zeros 
that appear in the denominators of Nekrasov's partition functions if we apply the AGT prescription 
without modification. They indicate that we have included zero-norm states among the states that 
flow in the channels of the conformal blocks. 

\subsubsection{The zeros in the numerators}
These zeros are due to the vanishing of the coupling of the zero-norm states and all other states.
In \cite{bershtein.foda}, it was shown, in the case of Virasoro minimal models, that for every 
zero in a denominator, there is a zero in the numerator, but the reverse is not true. In other 
words, the set of terms that contain a zero in the denominator is a proper subset of the set of 
terms that contain a zero in the numerator. 
We have not shown that this is the case here, since we do not need it for the purposes of this 
paper, but it is a straightforward, albeit tedious exercise to show that this is the case.
This ensures that one never has terms in the form of a finite number divided by zero, that are 
strictly infinite, but that one has ill-defined terms in the form of zero divided by zero.

\subsubsection{Resolving the ambiguities}
Assuming that for every zero in a denominator, there is a higher-degree zero in the numerator, 
one way to avoid 
the ill-defined expressions described above is to deform the conformal field theory away from 
minimality, such that all denominators become non-zero, then carefully prove that the minimal 
limit exists, presumably by showing that the numerators are always zero, or always vanish faster 
than the denominators. We are able to do this in simple, specific examples, but we have no proof 
that this is always the case. In this work, we pursue a different approach.

\subsection{Modifying $\cW_N$ AGT to apply to minimal model conformal blocks}
In this work, as in \cite{alkalaev.belavin, bershtein.foda}, we avoid the ill-defined expressions 
by restricting the summations over the $N$-partitions that appear in the sum (\ref{agt.gen}). 
We start from the original expression for $\cB^{\, p, \, \pp, \cH}_{N, n}$ in terms of sums of 
type (\ref{agt.gen}), each of which is a product of building blocks $Z^{\i}_{bb}$. 
We characterize the singularities in $Z^{\i}_{bb}$ that lead to ill-defined expressions, and 
eliminate these zero-norm states by restricting the $N$-partitions that appear in (\ref{agt.gen}) 
to $N$-partitions $\vec{Y}=\{Y_1, \cdots, Y_N\}$, that satisfy the conditions $N$-Burge conditions,
which we recall here, 

\begin{equation}
\label{N.Burge.conditions.repeated}
\boxed{
Y_{i, \, \rrho}-Y_{i+1, \, \rrho+s_i-1}\geq -r_i+1
}
\end{equation}

\noindent where $Y_{i,  \, \rrho}$ is the $\rrho$-row of $Y_i, i = 1, \cdots, N$, $r_{\i}$ and $s_{\i}$ 
are parameters that characterise the $\cW_N$ irreducible highest weight module that flows in a channel 
in a minimal model conformal block, and satisfy equation {\bf \ref{sum.conditions}}, and $Y_{N+1} = Y_1$.
{\it Note that we characterise the Young diagrams that do not lead to zeros, and only these. In other 
words, the Burge conditions are sufficient and necessary conditions for the procedure to work.} This 
is the reason why in section {\bf \ref{wn.characters}}, we obtain the correct character expressions.

For $N=2$, the {\it $N$-Burge partitions} were introduced in \cite{burge}, and further studied in 
\cite{foda.lee.welsh}. 
They appeared in full generality in \cite{gessel.krattenthaler}, and in the form used in this work 
in \cite{feigin.feigin.jimbo.miwa.mukhin.02}. We have shown that when used 
to restrict AGT to compute $\cB^{\, p, \, \pp, \, \cH}_{N, n}$, we obtain the expressions which we 
recall here,
\begin{equation}
\label{restricted.sum.repeated}
\boxed{
\cB^{\, p, \, \pp, \, \cH}_{N, n} =\sum^{\prime}_{\vec Y^1, \cdots, \vec Y^n}
\prod_{\i = 1}^{n+1} q_{\i}^{| \vec Y^{\i } |} 
Z_{bb}^{\i} 
\ll \vP_{\vec r_{\i - 1} \ \vec s_{\i - 1}}, \vec Y^{\i - 1}\ | \ a_{m_{\i} n_{\i} } \ | 
  \ \vP_{\vec r_{\i    } \ \vec s_{\i    }}, \vec Y^{\i} \rr
}
\end{equation}
\noindent where $\sum^{\prime}$ indicates that the sum is restricted to $N$-partitions that 
satisfy the $N$-Burge conditions (\ref{N.Burge.conditions.repeated}), which are well-defined 
expressions that we identify with $\cW_N$ minimal model conformal blocks, times Heisenberg 
factors. We check our identification in a non-trivial case, and show that it produces the 
correct 0-point conformal blocks on the torus, in specific cases.

\subsection{Related works}

\noindent {\bf 1.} In \cite{santachiara.tanzini}, Santachiara and Tanzini apply AGT to compute conformal blocks of 
$\{r, s\} = \{1, 2\}$ and $\{2, 1\}$ vertex operators in Virasoro minimal models. The ill-defined 
expressions were circumvented using an analytic continuation scheme that was tested to low orders 
in the combinatorial expansion of the instanton partition functions. 

If one can extend the analytic continuation scheme used in \cite{santachiara.tanzini} to the full 
instanton partition functions of the most general conformal blocks, and obtain the same result as 
in the present work, then this would amount to a proof that the proposed modified AGT expression 
for $\cB^{\, p, \, \pp, \, \cH}_{N, n}$ in equation (\ref{restricted.sum}) is indeed the required 
minimal model conformal block up to a Heisenberg factor.

\noindent {\bf 2.} In \cite{estienne.pasquier.santachiara.serban}, Estienne, Pasquier, Santachiara 
and Serban study conformal blocks of vertex operators such that 
$r_1 = 2$, and $r_i = 1$, $i = 2, \cdots, N-1$, and $s_i = 1$, $i = 1, \cdots, N$, or  
$r_i = 1$, $i = 1, \cdots, N$, $s_1 = 2$, and $s_i = 1$, $i = 2, \cdots, N-1$.  
in $\cW_N^{\, p, \pp} \oplus \cH$ minimal models.
From the null-state conditions of these vertex operators, Estienne {\it et al.} show that these 
specific conformal blocks are labeled by $N$-partitions that satisfy specific conditions. 
While the notation used in \cite{estienne.pasquier.santachiara.serban} is different from that 
in this work, one can check, in simple cases, that their $N$-partitions are equivalent to those 
that appear in this work. 

\noindent {\bf 3.} In \cite{fucito.morales.poghossian}, Fucito, Morales and Poghossian 
show that $\cN \! = \! 2$ supersymmetric Yang-Mills gauge theories on the squashed $S^4$, with 
rational deformation parameters, are dual to Virasoro minimal models. Ill-defined expressions 
are handled using a deformation scheme, akin to that used in \cite{santachiara.tanzini}, and 
rested to low orders in the combinatorial expansion of the instanton partition functions.

\noindent {\bf 4.} In \cite{alkalaev.belavin, bershtein.foda}, as outlined in section 
{\bf \ref{introduction}}, 
Virasoro minimal model conformal blocks are derived, via a modification of the AGT prescription, 
from the instanton partition functions of $\cN \! = \! 2$ supersymmetric $U(2)$ quiver gauge 
theories. In \cite{foda.wu.01}, the building block of the instanton partition functions that 
appeared in \cite{alkalaev.belavin, bershtein.foda} is derived by gluing four copies of refined 
topological vertices \cite{ikv} to form the partition function of a {\it strip geometry}, then 
choosing the gluing parameters and the partitions that label the unglued external legs of the 
strip appropriately. In \cite{foda.wu.02}, the building block of the instanton partition function 
that is used in the present work to generate $\cW_N$ minimal model conformal block is derived 
from refined topological vertices, using vertex operator methods, along the lines of 
\cite{okounkov.reshetikhin.02, okounkov.reshetikhin.vafa}.

\noindent {\bf 5.} In \cite{fukuda.nakamura.matsuo.zhu}, Fukuda, Nakamura, Matsup and Zhu studied 
the representation theory of $SH^c$, the central extension of the degenerate double affine Hecke 
algebra \cite{schiffmann.vasserot, kanno.matsuo.zhang} in the context of the minimal $\cW_N$ models. 
They found, among other results, that the states are labelled by $N$-partitions that satisfy the 
$N$-Burge conditions discussed in this work.

\subsection{Open problems}

\noindent {\bf 1.} This work may be regarded as an attempt to understand $\cW_N$ minimal model conformal 
blocks, that is, expectation values of degenerate $\cW_N$ vertex operators, in 2D conformal field 
theories, in terms of instanton partition functions in 4D $\cN \! = 2 \!$ supersymmetric gauge theories. 
However, the meaning of the choice of gauge theory parameters that lead to minimal $\cW_N$ theories, as 
well the interpretation of the $N$-Burge conditions at the level of 4D gauge theories remains unclear. 
One way to address these issues is to use the interpretation of the 2D degenerate vertex operators in 
terms of 4D surface operators along the lines of 
\cite{alday.gaiotto.gukov.tachikawa.verlinde, drukker.gomis.okuda.teschner}, where 
the expectation value of an elementary surface operator, in a 4D $\cN \! = 2 \!$ supersymmetric gauge 
theory, is shown to be equal to the expectation value of vertex operators in a 2D Liouville conformal 
field theory, in the presence of a degenerate vertex operator of type $\cO_{2, 1} (z)$. 

The literature on the 2D degenerate vertex operator/4D surface operator connection is extensive, 
and beyond the limited scope of this work, see \cite{gukov} for a review. But we expect that the 
adaptation of 2D degenerate operator/4D surface operator connection to AGT in the context of minimal 
models will help clarify the issues outlined above. 

\noindent {\bf 2.} In the present work, we have restricted our attention to $\cW_N$ conformal blocks 
that 
satisfy the FLW conditions of subsection {\bf \ref{flw.condition}}. In \cite{gomis.lefloch}, Gomis and 
LeFloch propose that one can obtain $\cW_N$ Toda conformal blocks that are expectation values of vertex 
operators that include degenerate $\cW_N$ vertex operators that do not satisfy the FLW conditions, in 
addition to non-degenerate vertex operators, and interpret the degenerate operators at the gauge theory 
level as surface operator insertions. More precisely, the proposal of Gomis and LeFloch is that one can
obtain the degenerate vertex operator insertions that do not satisfy the FLW conditions by starting from
vertex operators insertions that satisfy the conditions, then bringing the latter together in a form 
of operator product expansion. While formally plausible, it is not clear to us at this stage whether 
the proposal of Gomis and LeFloch leads to tractable results along the lines of the $\cW_N$ AGT results 
presented in this work. 

\section*{Acknowledgements}
OF thanks M Bershtein for collaboration on \cite{bershtein.foda}, ideas of which were used 
in this work, J F Morales, E Pomoni and R Poghossian for discussions on their works and related topics. 
OF also acknowledges the excellent hospitality and financial support of the Galileo Galilei 
Institute, University of Florence, and particularly of the scientific program 
{\it \lq\lq Statistical Mechanics, Integrability and Combinatorics\rq\rq}.
RS thanks G Bonelli and A Tanzini for discussions. 
All authors thank B Estienne and B LeFloch for discussions, and the Institut Henri Poincare, 
Paris, where this work was completed, for excellent hospitality and financial support. 
We also thank the referee for pointing out \cite{alday.gaiotto.gukov.tachikawa.verlinde}
to our attention.
VB is supported by the Russian Science Foundation under the grant 14-12-01383.
OF is supported by the Australian Research Council under project DP140103104.


\begin{thebibliography}{99}

%01
\bibitem{agt}
L F Alday, D Gaiotto and Y Tachikawa,
{\it Liouville Correlation Functions from Four-dimensional Gauge Theories},
Letters in Mathematical Physics {\bf 91.2} (2010) 167--197.
{\tt arXiv:0906.3219} 

%02
\bibitem{fateev.litvinov.2009}
V A Fateev and A V Litvinov, 
{\it On AGT conjecture},
Journal of High Energy Physics {\bf 1002} (2010) 014,
{\tt arXiv:0912.0504}
%
%03
\bibitem{mironov.morozov.shakirov.01}
A Mironov, A Morozov and Sh Shakirov, 
{\it Towards a proof of AGT conjecture by methods of matrix models},
International Journal of Modern Physics {\bf A 27.01} (2012) 1230001
{\tt arXiv:1011.5629}
%
%04
\bibitem{mironov.morozov.shakirov.02}
A Mironov, A Morozov and Sh Shakirov, 
{\it A direct proof of AGT conjecture at beta = 1} 
Journal of High Energy Physics {\bf 1102} (2011) 067
{\tt arXiv:1012:3137}
%
%05
\bibitem{hadasz.jaskolski.suchanek.2010}
L Hadasz, Z Jaskolski and P Suchanek,
{\it Proving the AGT relation for $N_f = 0, 1, 2$ antifundamentals},
Journal of High Energy Physics {\bf 1006} (2010) 046,
{\tt arXiv:1004.1841} 
%
%06
\bibitem{schiffmann.vasserot}
O Schiffmann and E Vasserot,
{\it Cherednik algebras, $W$ algebras and the equivariant cohomology of the moduli space of instantons on $A^2$},
Publications mathematiques de l'IHES {\bf 118}, no. 1 (2013) 213--342.
{\tt arXiv:1202.2756} 
%
%07
\bibitem{alba} 
V A Alba, V A Fateev, A V Litvinov and G M Tarnopolskiy,
{\it On combinatorial expansion of the conformal blocks arising from AGT conjecture},
Letters in Mathematical Physics {\bf 98} (2011) 33--64
{\tt arXiv:1012.1312} 
%
%08
\bibitem{bouwknegt.schoutens.review}
P Bouwknegt and K Schoutens, 
{\it $\cW$-symmetry in conformal field theory},
Physics Reports {\bf 223.4} (1993) 183--276.
{\tt hep-th/9210010}
%
%09
\bibitem{wyllard} 
N Wyllard,
{\it $A_{(N-1)}$ conformal Toda field theory correlation functions from conformal 
$\cN = 2$ $SU(N)$ quiver gauge theories},
Journal of High Energy Physics {\bf 2009.11} (2009) 002.
{\tt  arXiv:0907.2189}
%
%10
\bibitem{mironov.morozov}
A Mironov and A Morozov, 
{\it On AGT relation in the case of $U(3)$}, 
Nuclear Physics {\bf B 825} (2010) 1--37.
{\tt arXiv:0908.2569}
%
%11
\bibitem{fateev.litvinov.2011}
V A Fateev and A V Litvinov,
{\it Integrable structure, $\cW$-symmetry and AGT relation}, 
Journal of High Energy Physics {\bf 1201} (2012) 051
{\tt arXiv:1109.4042 [hep-th]}
%
%12
\bibitem{fateev.litvinov.2008}
V A Fateev and A V Litvinov,
{\it Correlation functions in conformal Toda field theory. II.},
Journal of High Energy Physics {\bf 0701} (2009) 033
{\tt [arXiv:0810.3020 [hep-th]}
%
%13
\bibitem{alkalaev.belavin} 
K B Alkalaev and V A Belavin, 
{\it Conformal blocks of $\cW_N$ Minimal Models and AGT correspondence},
Journal Of High Energy Physics {\bf 2014}, no. 7 (2014) 1--16.
{\tt arXiv:1404.7094} 
%
%14
\bibitem{bershtein.foda}
M Bershtein and O Foda,
{\it AGT, Burge pairs and minimal models},  
Journal of High Energy Physics {\bf 2014}, no. 6 (2014) 1--29.
{\tt arXiv:1404.7075 [hep-th]}
%
%15
\bibitem{burge}
W H Burge, 
{\it Restricted partition pairs}, 
J of Comb Th {\bf A 63.2} (1993) 210--222.
%
%16
\bibitem{foda.lee.welsh}
O Foda, K S M Lee and T A Welsh,
{\it A Burge tree of Virasoro-type polynomial identities},
International Journal of Modern Physics {\bf A 13}, no. 29 (1998) 4967--5012.
{\tt  arXiv:q-alg/9710025}
%
%17
\bibitem{gessel.krattenthaler}
I Gessel and Ch Krattenthaler,
{\it Cylindric partitions},
Transactions of the American Mathematical Society {\bf 349}, no. 2 (1997) 429-479.
%
%18
\bibitem{feigin.feigin.jimbo.miwa.mukhin.02}
B Feigin, E Feigin, M Jimbo, T Miwa and E Mukhin, 
{\it Quantum continuous $gl_\infty$: Semi-infinite construction of representations},
Kyoto Journal of Mathematics {\bf 51}, no. 2 (2011) 337--364.
{\tt arXiv:1002.3100}
%
%19
\bibitem{belavin.polyakov.zamolodchikov}
A A Belavin, A M Polyakov and A B Zamolodchikov, 
{\it Infinite conformal symmetry in two-dimensional quantum field theory},
Nucl Phys {\bf B 241}, no. 2 (1984) 333--380.
%
%20
\bibitem{zamolodchikov.w.algebra} 
A B Zamolodchikov,
{\it Infinite additional symmetries in 2-dimensional conformal quantum field theory},  
Teor Mat Phys {\bf 65} (1985) 1205--1213.
%
%21
\bibitem{iles.watts}
N J Iles and G M T Watts, 
{\it Modular properties of characters of the $\cW_3$ algebra},
{\tt arXiv:1411.4039} 
%
%22
\bibitem{blumenhagen.1991} 
R Blumenhagen, M Flohr, A Kliem, W Nahm, A Recknagel, and R Varnhagen, 
{\it $\cW$-algebras with two and three generators},
Nuclear Physics {\bf B 361}, no. 1 (1991) 255--289.
%
%23
\bibitem{kausch.watts.1991}
H G Kausch, and G M T Watts, 
{\it A study of $\cW$-algebras using Jacobi identities},
Nuclear Physics {\bf B 354} (1991) 740--768.
%
%24
\bibitem{carlsson.okounkov}
E Carlsson and A Okounkov,
{\it Exts and vertex operators},
{\tt arXiv:0801.2565}
%
%25
\bibitem{nekrasov}
N A Nekrasov,
{\it Seiberg-Witten prepotential from instanton counting},
Advances in Theoretical and Mathematical Physics {\bf 7} (2004) 831--864. 
{\tt hep-th/0206161}
%
%26
\bibitem{kanno.matsuo.zhang}
A Kanno, Y Matsuo and H Zhang, 
{\it Extended conformal symmetry and recursion formulae for Nekrasov partition function}, 
Journal of High Energy Physics {\bf 2013}, no. 8 (2013) 1--27.
{\tt arXiv:1306.1523}
%
%27
\bibitem{fateev.litvinov.2007}
V A Fateev and A V Litvinov,
{\it Correlation functions in conformal Toda field theory. I.},
Journal of High Energy Physics {\bf 0711} (2007) 002  
{\tt arXiv:0709.3806}
%
%28
\bibitem{zam.zam}
A Zamolodchikov and Al Zamolodchikov, 
{\it Conformal bootstrap in Liouville field theory}, 
Nuclear Physics {\bf B 477}, no. 2 (1996) 577-605.
{\tt hep-th/9506136}
%
%29
\bibitem{frenkel.kac.wakimoto} 
E Frenkel, V Kac and M Wakimoto, 
{\it Characters and fusion rules for $\cW$-algebras via quantized Drinfel'd-Sokolov reduction}, 
Communications in Mathematical Physics {\bf 147} no. 2 (1992) 295--328
%
%30
\bibitem{feigin.feigin.jimbo.miwa.mukhin.01}
B Feigin, E Feigin, M Jimbo, T Miwa and E Mukhin, 
{\it Quantum continuous $\mathfrak {gl}_{\infty}$: Tensor products of Fock modules and ${\cW}_n$-characters},
Kyoto Journal of Mathematics {\bf 51}, no. 2 (2011) 365--392.
{\tt arXiv:1002.3100}
%
%31
\bibitem{santachiara.tanzini}
R Santachiara and A Tanzini,
{\it Moore-Read fractional quantum Hall wave functions and $SU(2)$ quiver gauge theories},
Physical Review {\bf D 82}, no. 12 (2010) 126006.
{\tt  arXiv:1002.5017}
%
%32
\bibitem{estienne.pasquier.santachiara.serban}
B Estienne, V Pasquier, R Santachiara and D Serban,
{\it Conformal blocks in Virasoro and W theories: Duality and the Calogero-Sutherland model},
Nuclear Physics {\bf B 860}, no. 3 (2012) 377--420.
{\tt  arXiv:1110.1101}
%
%33
%
\bibitem{fucito.morales.poghossian}
F Fucito, J F Morales and R Poghossian, 
{\it Wilson Loops and Chiral Correlators on Squashed Sphere},
{\tt arXiv:1507.05426} 
%
%34
%
\bibitem{foda.wu.01}
O Foda and J-F Wu,
{\it From topological strings to minimal models},
Journal of High Energy Physics {\bf 2015} (2015).
{\tt arxiv:1504.01925}
%
%35
\bibitem{ikv}
A Iqbal, C Kozcaz and C Vafa, 
{\it The refined topological vertex}, 
Journal of High Energy Physics 10 (2009) 069, 
{\tt hep-th/0701156}
%
%36
%
\bibitem{foda.wu.02}
O Foda and J-F Wu,
in preparation.
%
%37
%
\bibitem{okounkov.reshetikhin.02}
A Okounkov and N Reshetikhin,
{\it Random skew plane partitions and the Pearcey process},
Communications in mathematical physics {\bf 269.3} (2007) 571-609,
{\tt arxiv.math:0503508}
%
%38
\bibitem{okounkov.reshetikhin.vafa}
A Okounkov, N Reshetikhin and C Vafa, 
{\it Quantum Calabi-Yau and classical crystals},
in 
{\it The unity of Mathematics: In Honor of the Ninetieth Birthday of I M Gelfand},
Progress in Mathematics {\bf 244}, 597-618,
P Etingof, V S Retakh and I M Singer, Editors, 
Birkhauser (2006), 
{\tt {\bf ISBN-13:} 978-0817640767}
{\tt hep-th/0309208}
%
%39
\bibitem{fukuda.nakamura.matsuo.zhu}
M Fukuda, S Nakamura, Y Matsuo and R-D Zhu,
{\it $SH^c$ realization of minimal model CFT: Triality, Poset and Burge conditions},
{\tt arXiv:1509.01000}
%
%40
\bibitem{alday.gaiotto.gukov.tachikawa.verlinde}
L F Alday, D Gaiotto, S Gukov, Y Tachikawa and H Verlinde,
{\it Loop and surface operators in $\cN \! = \! 2$ gauge theory and Liouville 
modular geometry},
Journal of High Energy Physics {\bf 2010}, no. 1 (2010) 1--50,
{\tt arXiv:0909.0945}
%
%41
\bibitem{drukker.gomis.okuda.teschner}
N Drukker, J Gomis, T Okuda, and J Teschner, 
{\it Gauge theory loop operators and Liouville theory},
Journal of High Energy Physics {\bf 2010}, no. 2 (2010) 1--62.
{\tt arXiv:0909.1105}
%
%42
\bibitem{gukov}
S Gukov,
{\it Surface Operators},
{\tt arXiv:1412.7127}
%
%43
\bibitem{gomis.lefloch}
J Gomis and B Le Floch, 
{\it M2-brane surface operators and gauge theory dualities in Toda},
{\tt arXiv:1407.1852} 
%
\end{thebibliography}
\end{document}